\documentclass[12pt]{article}
\usepackage{graphicx}
\usepackage{amssymb}
\begin{document}
\def\uu{\langle \bar u u \rangle}
\def\dd{\langle \bar d d \rangle}
\def\ss{\langle \bar s s \rangle}
\def\qq{\langle \bar q q \rangle}
\def\figt#1#2#3{
	\begin{figure}
	$\left. \right.$
	\vspace*{-2cm}
	\begin{center}
	\includegraphics[width=10cm]{#1}
	\end{center}
	\vspace*{-0.2cm}
	\caption{#3}
	\label{#2}
	\end{figure}
		}
\def\figb#1#2#3{
	\begin{figure}
	$\left. \right.$
	\vspace*{-1cm}
	\begin{center}
	\includegraphics[width=10cm]{#1}
	\end{center}
	\vspace*{-0.2cm}
	\caption{#3}
	\label{#2}
	\end{figure}
		}

\title{
Radiative Decays of Decuplet to Octet Baryons in Light Cone QCD
}
\author{
T. M. Aliev\thanks{taliev@metu.edu.tr}~\footnote{permanent address:Institute of Physics,Baku,Azerbaijan} \\
{\small Physics Dept., Middle East Technical Univ. 06531 Ankara, Turkey} \\
\and
A. Ozpineci\thanks{ozpineci@ictp.trieste.it}~\thanks{present adress: INFN,Sezione di Bari, Bari, Italy} \\
{\small HE Section, The Abdus Salam ICTP, I-34100, Trieste, Italy}
}
\begin{titlepage}
\maketitle
\thispagestyle{empty}
\begin{abstract}
The radiative decays of  decuplet to octet baryons are analyzed within the
light cone QCD sum rules framework.The electromagnetic transition form factors for
these decays are calculated at $q^2=0$ up to twist four accuracy for photon
wave functions as well as including first order strange quark mass
corrections. A comparison of our results with predictions of lattice
theory and existing experimental data is presented.

\end{abstract}
\end{titlepage}

\section{Introduction}

The study of the electromagnetic transitions of the decuplet to octet baryons is an important 
issue for understanding of internal structure of baryons. 
Spin parity selection rules allow for magnetic dipole ($M1$),
electric quadrapole ($E2$) and Coulomb quadrapole ($C2$) moments     for these decays. 
These transitions can also give essential information about the wave function 
of the lowest lying baryons. For example in the $\Delta \rightarrow N \gamma^*$
transition if initial and final baryons' wave functions are spherically symmetric,
then the $E2$ and $C2$ amplitude must vanish. The results of recent photo production
experiments on nucleon at Bates \cite{R1} and Jefferson Lab \cite{R2} shows that
these amplitudes are likely to be non zero. This result is an indication that these
decays can contain many mysteries. For this reason, these decays must be carefully
and completely studied both theoretically and experimentally.

The electromagnetic decays of baryons constitute an
important class of decays for the determination of fundamental parameters of
hadrons. For extracting these parameters, information about the
non perturbative region of QCD is required. Therefore a reliable non-perturbative
approach is needed. Among non-perturbative approaches, QCD sum rules \cite{R3} is
more predictive in studying properties of hadrons.

QCD sum rules is based on the first principles of QCD. In this method, measurable
quantities of hadrons in experiments are connected with QCD parameters, where
hadrons are represented by corresponding interpolating quark currents taken at large
virtualities and correlator of these quark currents is introduced. The main idea is
to calculate this correlator with the help of operator product expansion (OPE) in
the framework of QCD in the large Euclidean domain and represent this same correlator in terms of
hadronic parameters in the other kinematical region. 
The sum rule for corresponding physical quantity is obtained by matching
two representations of the correlator.

In the present work, we calculate the baryon decuplet to octet transition form
factors in the framework of an alternative approach to the traditional QCD sum
rules, i.e. light cone QCD sum rules method. This method is based on the OPE near
the light cone, which is an expansion over the twist of the operators rather than
dimensions as is the case in the traditional QCD sum rules. The non-perturbative dynamics encoded
in the light cone wave functions determines the matrix elements of the non-local
operators between vacuum and corresponding particle states (more about his method
can be found in \cite{R4,R5}).

It should be mentions that the electromagnetic baryon decuplet to octet transitions
has been investigated in chiral perturbation theory \cite{R6,R7,R8,R9}, (partially) 
quenched chiral perturbation theory \cite{R10}, in lattice simulations of quenched QCD in \cite{leinweber},
in the large $N_c$ limit of QCD
\cite{R11}. In \cite{R12} the transition form factors for $\Delta \rightarrow N
\gamma$ decay is calculated within the traditional QCD sum rules using external
field method. This $\Delta \rightarrow N \gamma$ decay in QCD light cone QCD sum
rules is investigates in \cite{R13}.

The plan of this work is as follows. In section 2, 
we consider a generic correlator function,which yields light cone
sum rules for decuplet to octet baryon transition form factors.
Then this correlation
function is calculated up to twist 4 accuracy for the photon wave functions 
including first order strange quark
mass corrections. In this section we obtain sum rules for 
form factors of decuplet to octet baryons electromagnetic transition. In section 3, we present our numerical
results and comparison with other approaches is presented.

\section{Sum Rules for Electromagnetic Transition Moments for the Decuplet to
Octet Baryon Transitions}
In this section we calculate the baryon decuplet to octet electromagnetic transition
form factors. We start our calculation by considering the following correlator
function:
\begin{eqnarray}
T_\mu(p,q) = i \int d^4 x e^{ipx} \langle \gamma(q) \vert 
{\cal T} \left\{ \eta_{\cal O}(x) \bar \eta_{{\cal D} \mu} (0) \right\} 
\vert 0 \rangle
\label{eq1}
\end{eqnarray}
where $\eta_{\cal O}$ and $\eta_{\cal D}$ are generic octet and decuplet
interpolating quark currents respectively. We will consider the following baryon
decuplet to octet electromagnetic transitions
\begin{eqnarray}
\Sigma^{*+} &\rightarrow& \Sigma^+ \gamma \nonumber \\
\Sigma^{*0} &\rightarrow& \Sigma^0 \gamma \nonumber \\
\Sigma^{*0} &\rightarrow& \Lambda \gamma \nonumber \\
\Sigma^{*-} &\rightarrow& \Sigma^- \gamma \nonumber \\
\Xi^{*0} &\rightarrow& \Xi^0 \gamma \nonumber \\
\Xi^{*-} &\rightarrow& \Xi^- \gamma 
\end{eqnarray}

First, the
phenomenological part of the correlator function can be calculated by
inserting a complete set of hadronic states to it.
Saturating the
correlator (Eq. (\ref{eq1})) by ground state baryons we get:
\begin{eqnarray}
T_\mu(p,q) = 
\frac{\langle 0 \vert \eta_{\cal O} \vert \frac12(p) \rangle}{p^2 -m_{\cal O}^2} 
\langle \frac12 \vert \frac32 \rangle_\gamma
\frac{\langle \frac32(p+q) \vert \eta_{{\cal D}\mu} \vert 0 \rangle}{(p+q)^2-M_{\cal D}^2} + \cdots
\label{eq2}
\end{eqnarray} 
where $\vert \frac12 (p) \rangle$ and $\vert \frac32 (p) \rangle$ denote octet and decuplet baryons with momentum $p$, respectively, 
$m_{\cal O}$ and $M_{\cal D}$ denote the masses of the octet and decuplet baryons respectively, and $\cdots$
stand for the contributions of the higher states and the continuum.
The matrix elements $\langle 0 \vert \eta_{\cal O} \vert \frac12(p) \rangle$ and
$\langle \frac32(p+q) \vert \eta_{{\cal D}\mu} \vert 0 \rangle$ are determined as:
\begin{eqnarray}
\langle 0 \vert \eta_{\cal O} \vert \frac12(p) \rangle &=& \lambda_{\cal O} u(p,s)
\nonumber \\
\langle \frac32(p+q) \vert \eta_{{\cal D}\mu} \vert 0 \rangle &=& \lambda_{\cal D}
u_\mu(p+q,s')
\label{eq4}
\end{eqnarray}
where $\lambda_{\cal O}$ and $\lambda_{\cal D}$ are the residues and $u_\mu$ is the
Rarita-Schwinger spinor and $s$ and $s'$ are the four spin vectors of the octet and
decuplet baryons respectively. The electromagnetic vertex of the decuplet to octet
transition can be parameterized in terms of three form factors in the following way
\cite{R15,R16}
\begin{eqnarray} 
\langle \frac12 \vert \frac32 \rangle_\gamma &=&
e u(p,s) 
\left\{  G_1 \left( q_\rho \not\!\varepsilon - \varepsilon_\rho \not\!q \right) \gamma_5 \right.
\nonumber \\
&+& G_2 \left( (P\varepsilon) q_\rho - (Pq) \varepsilon_\rho \right) \gamma_5
\nonumber \\ 
&+& \left. G_3 \left( (q\varepsilon) q_\rho - q^2 \varepsilon_\rho \right) \gamma_5 
\right\} u^\rho(p+q)
\label{eq5}
\end{eqnarray}
where $P = \frac{1}{2} \left(p + (p+q)\right)$ and $\varepsilon$ is the photon polarization vector. 
Since in our case, photon is real, $G_3$ does not give any
contribution to the considered decays and we need to know the values of the form
factors $G_1$ and $G_2$ only at $q^2=0$.

For the experimental analysis, it is desirable to use such form factors which
describe physical transition. This would correspond to a definite multipole or
helicity transitions in a given reference frame. Linear combinations of the the
form factors in Eq. (\ref{eq5}) give magnetic dipole, $G_M$, electric quadrapole,
$G_E$, and Coulomb quadrapole, $G_C$, form factors. The relations between two set of
form factors (for completeness, we present the relation in general form when $q^2
\neq 0$. Our case corresponds to $q^2=0$):
\begin{eqnarray}
G_M &=& \left[ \left( \left(3 M_{\cal D}+m_{\cal O} \right) (M_{\cal D}+m_{\cal O})-q^2 \right) \frac{G_1}{M_{\cal D}} 
\right. \nonumber \\ && \left.
+\left( M_{\cal D}^2-m_{\cal O}^2 \right) G_2 + 2 q^2 G_3 \right] \frac{m_{\cal O}}{3 (M_{\cal D}+m_{\cal O})}
\nonumber \\
G_E &=& \left[ \left(M_{\cal D}^2 - m_{\cal O}^2 + q^2 \right) \frac{G_1}{M_{\cal D}} +
\left( M_{\cal D}^2-m_{\cal O}^2 \right) G_2 + 2 q^2 G_3 \right] \frac{m_{\cal O}}{3 (M_{\cal D}+m_{\cal O})}
\nonumber \\
G_C &=& \left[ 2 M_{\cal D} G_1 +
\frac12 \left(3 M_{\cal D}^2 + m_{\cal O}^2 - q^2 \right) G_2 
\right. \nonumber \\ && \left. 
+ \left( M_{\cal D}^2 - m_{\cal O}^2 + q^2 \right) G_3
\right] \frac{2 m_{\cal O}}{3 (M_{\cal D}+m_{\cal O})}
\label{eq6}
\end{eqnarray}
Considering the expression for $G_E$, one sees that in the case we are interested in, i.e. $q^2=0$, $G_E$ is
proportional to $M_{\cal D} -m_{\cal O}$. 
As this quantity is very small ($(M_{\cal D}-m_{\cal O})/(M_{\cal D}+m_{\cal O})$ is $\sim 15\%$ for the $\Delta \rightarrow N$
transitions and is $\sim 5\%$ for the remaining transitions), small uncertainties in $M_{\cal D}$ or $m_{\cal O}$ leads to big uncertainties in the
predictions for $G_E$. Although the sum rule predictions for the masses of the octet and decuplet baryons agree within errors with the experimental values,
the error get amplified in the prediction for $G_E$. Thus it makes a big difference weather one uses the experimental value or the sum rule 
prediction for the values of $M_{\cal D}$ and $m_{\cal O}$. 
To give a quantitative idea about this uncertainty, if one considers the $\Delta \rightarrow N$ transitions,
the experimental value of the mass difference of the $\Delta$ and the nucleon is $M^{exp}_\Delta-m^{exp}_N = 294~MeV$, where as the mass 
prediction from the mass sum rules for the mass of the $\Delta$ can be as small as $M_\Delta=1.06~GeV$ \cite{fxlee15} and the prediction for the mass
of the nucleon can be as big as $m_N = 1.17~GeV$ \cite{fxlee2}. Thus one sees that, in the extreme cases, if one uses the predictions of the
mass sum rules for the masses of the octet and decuplet baryons, even the sign of $G_E$ might change. In our work, we use the experimental values for
the masses of the octet and decuplet baryons. But our predictions on $G_E$ should be considered as order of magnitude estimates.

In calculations, summation over spins of the Rarita-Schwinger spin vector is done 
using the relation:
\begin{eqnarray}
\sum_s u_\alpha(p,s) \bar u_\beta(p,s) = - \left( \not\!p+M_{\cal D} \right)  \left\{
g_{\alpha \beta} - \frac13 \gamma_\alpha \gamma_\beta - 
\frac{2 p_\alpha p_\beta}{3 M_{\cal D}^2} + \frac{ p_\alpha \gamma_\beta - p_\beta
\gamma_\alpha}{3 M_{\cal D}} \right\}
\label{eq7}
\end{eqnarray}

Using Eqs. (\ref{eq2}-\ref{eq5}) and (\ref{eq7}),
we see that correlator has numerous tensor structures not all
of which are independent. The dependence can be removed by ordering gamma matrices
in a specific order. For this purpose, the ordering $\not\!\varepsilon \not\!q \not\!p
\gamma_\mu$ is chosen. With this ordering, the correlator function becomes:
\begin{eqnarray}
T_\mu &=& e \lambda_{\cal O} \lambda_{\cal D} \frac{1}{p^2-m_{\cal O}^2} \frac{1}{(p+q)^2-M_{\cal D}^2} \left[ \right. 
\nonumber \\ &&
\left[ \varepsilon_\mu (pq) - (\varepsilon p) q_\mu \right] \left\{
-2 G_1 M_{\cal D} - G_2 M_{\cal D} m_{\cal O} + G_2 (p+q)^2 \right.
\nonumber \\ &&
+ \left. \left[ 2 G_1 + G_2(m_{\cal O}-M_{\cal D}) \right] \not\!p + m_{\cal O} G_2 \not\!q - G_2 \not\!q\not\!p \right\} \gamma_5 
\nonumber \\ && 
+ \left[ q_\mu \not\!\varepsilon - \varepsilon_\mu \not\!q \right] \left\{ G_1 (p^2 + M_{\cal D} m_{\cal O}) - 
G_1 (M_{\cal D}+m_{\cal O}) \not\!p \right\} \gamma_5
\nonumber \\ &&
+ 2 G_1 \left[ \not\!\varepsilon (pq) - \not\!q (\varepsilon p) \right] q_\mu \gamma_5
\nonumber \\ &&
- G_1 \not\!\varepsilon \not\!q \left\{ m + \not\!p \right\} q_\mu \gamma_5
\nonumber \\ &&
\left. \mbox{other structures with $\gamma_\mu$ at the end or which are proportional to $(p+q)_\mu$} \right]
\nonumber \\ &&
\label{eq9}
\end{eqnarray}

An essential reason for not considering the structures $\propto (p+q)_\mu$ and the structures which contain a $\gamma_\mu$ at the
end is as follows:
A spin-$\frac{3}{2}$ current can have a nonzero overlap with a spin-$\frac{1}{2}$ state:
\begin{eqnarray}
\langle 0 \vert \eta_{\frac{3}{2}\mu} \vert \frac{1}{2}(p) \rangle = 
\left( A' p_\mu + B' \gamma_\mu \right) \gamma_5 u(p)
\end{eqnarray} 
Using $\gamma^\mu \eta_{\frac{3}{2}\mu} = 0$ one can easily obtain that $B' = - \frac{A'
m}{4}$
Hence, in principle, spin-$\frac{1}{2}$ states can also give contribution to the
correlation function. But given the ordering, 
they contribute only to the structures which contain a $\gamma_\mu$ at the end or which are
proportional to $(p+q)_\mu$. In other words, spin-$\frac{1}{2}$ particles do not contribute to the structures that
we have not omitted.

In order to obtain predictions for the values of $G_1$ and $G_2$, we need to choose two different 
structures to study. In deciding which structures to study, we have chosen structures which do not get contributions from the
contact terms after the Borel transformations(for a discussion of the contact terms see \cite{wyler}). 
In the exact $SU(3)_f$ limit, the contact terms vanish. 
But since we are interested also in the violations of $SU(3)$, the contact terms are in principle non-zero. 
In order to extract the physical interaction vertex, one has to subtract the contact terms from the correlation function. Among the structures in
Eq. (\ref{eq9}), the structures which receive contributions from the contact terms are $\varepsilon_\mu \gamma_5,~\varepsilon_\mu \not\!q \gamma_5,
~\varepsilon_\mu \not\!p \gamma_5 ,~ q_\mu \gamma_5,~q_\mu \not\!q \gamma_5,$ and  $q_\mu \not\!p\gamma_5$

We have studied all the possible pairs of structures. We obtained the best convergence for the pairs
$\not\!\epsilon \not\!p \gamma_5 q_\mu$ and $\not\!q \not\!p \gamma_5 (\varepsilon p) q_\mu$, or
$\gamma_5 (\varepsilon p)q_\mu$ and $\not\!q \not\!p \gamma_5 (\varepsilon p) q_\mu$.
Although the structure $\not\!q \not\!p \gamma_5 \varepsilon_\mu$ receives contributions from the contact terms, the contact terms vanish
after the Borel transformation.

The structures $\not\!\varepsilon \not\!p \gamma_5 q_\mu$, $\gamma_5 (\varepsilon p) q_\mu$, and
$\not\!q \not\!p \gamma_5 (\varepsilon p) q_\mu$ have the
coefficients:
\begin{eqnarray}
- e \lambda_{\cal O} \lambda_{\cal D} (M_{\cal D}+m_{\cal O}) \Sigma_6 &=& - e \lambda_{\cal O} \lambda_{\cal D} (M_{\cal D}+m_{\cal O}) G_1 \nonumber \\
e \lambda_{\cal O} \lambda_{\cal D} M_{\cal D} \Sigma_9 &=& e \lambda_{\cal O} \lambda_{\cal D} M_{\cal D} (2 G_1 + G_2 (m_{\cal O} - M_{\cal D}))
\nonumber \\ 
e \lambda_{\cal O} \lambda_{\cal D} \Sigma_{12} &=& e \lambda_{\cal O} \lambda_{\cal D} G_2
\label{sigma}
\end{eqnarray}
respectively, where
\begin{eqnarray}
\Sigma_6 &=& G_1 \nonumber \\ 
\Sigma_9 &=& 2 G_1 + G_2 (m_{\cal O} - M_{\cal D}) \nonumber \\ 
\Sigma_{12} &=& G_2
\label{invert}
\end{eqnarray} 
Once the sum rules for $\Sigma_i$ ($i=4,~6,~12$) are obtained, the sum rules for  $G_X$, ($X=1,~2,~E,~M$) are obtained using the
experimental masses of the octet and decuplet baryons through the relations obtained by inverting
Eqs. (\ref{eq6}) and (\ref{invert}):

In the deep Euclidean region where  $p^2 << 0$ and $(p+q)^2 << 0$, the correlation function
Eq. (\ref{eq1}) can be calculated using OPE. For this purpose, one needs the explicit forms of the
interpolating currents.
Explicit forms of the interpolating quark currents for the corresponding decuplet
members are as follows:
\begin{eqnarray}
\eta_\mu^{\Sigma^{*0}} &=& \sqrt\frac23 \epsilon^{abc}
\left[
\left( u^{aT} C \gamma_\mu d^b \right) s^c +
\left( d^{aT} C \gamma_\mu s^b \right) u^c +
\left( s^{aT} C \gamma_\mu u^b \right) d^c
\right] \nonumber \\
\eta_\mu^{\Sigma^{*+}} &=& \frac{1}{\sqrt2} \eta_\mu^{\Sigma^{*0}}(d \rightarrow u)
\nonumber \\
\eta_\mu^{\Sigma^{*-}} &=& \frac{1}{\sqrt2} \eta_\mu^{\Sigma^{*0}}(u \rightarrow d)
\nonumber \\
\eta_\mu^{\Delta^{*+}} &=&  \eta_\mu^{\Sigma^{*+}}(s \rightarrow d)
\nonumber \\ 
\eta_\mu^{\Delta^{*0}} &=& \eta_\mu^{\Sigma^{*-}}(s \rightarrow u)
\nonumber \\ 
\eta_\mu^{\Xi^{*0}} &=& \eta_\mu^{\Delta^{0}}(d \rightarrow s)
\nonumber \\
\eta_\mu^{\Xi^{*-}} &=& \eta_\mu^{\Xi^{*0}}(u \rightarrow d)
\end{eqnarray}
It is well known that the choice of the interpolating currents for the octet baryons is not unique. One can define two different currents with the
quantum numbers of the octet baryons. The most general current is a linear combination of these two currents. In our studies we chose:
\begin{eqnarray}
\eta^{\Sigma^0} &=& \sqrt\frac12 \epsilon^{abc} \left[
\left( u^{aT} C s^b \right) \gamma_5 d^c + t \left( u^{aT} C \gamma_5 s^b \right) d^c
- \left( s^{aT} C d^b \right) \gamma_5 u^c - t \left( s^{aT} C \gamma_5 d^b \right) u^c
\right]
\nonumber \\
\eta^{\Sigma^+} &=& \frac{1}{\sqrt2} \eta^{\Sigma^0}(d \rightarrow u)
\nonumber \\ 
\eta^{\Sigma^-} &=& \frac{1}{\sqrt2} \eta^{\Sigma^0}(u \rightarrow d)
\nonumber \\
\eta^p &=& \eta^{\Sigma^+}(s \rightarrow d)
\nonumber \\
\eta^n &=&  \eta^{\Sigma^-}(s \rightarrow u)
\nonumber \\
\eta^{\Xi^0} &=& \eta^{n}(d \rightarrow s)
\nonumber \\
\eta^{\Xi^-} &=&  \eta^{\Xi^0}(u \rightarrow s)
\nonumber \\
\eta^{\Lambda} &=& -\sqrt\frac16 \epsilon^{abc} \left[
2 \left( u^{aT} C d^b \right) \gamma_5 s^c + 2 t \left( u^{aT} C \gamma_5 d^b \right) s^c
+\left( u^{aT} C s^b \right) \gamma_5 d^c 
\right. \nonumber \\ 
&+& t\left. \left( u^{aT} C \gamma_5 s^b \right) d^c
+\left( s^{aT} C d^b \right) \gamma_5 u^c + t \left( s^{aT} C \gamma_5 d^b \right) u^c
\right]
\label{eq3}
\end{eqnarray}
where $t$ is an arbitrary parameter. The Ioffe current, generally used in the literature to study the properties
of the octet baryons, corresponds to the choice $t=-1$ for this parameter.  The current used in \cite{leinweber} corresponds to the limit
$t\rightarrow \infty$.
The physical quantities should be independent of the value of $t$. In our analysis, a region for the values of $t$ is found requiring that
the prediction should be independent of the value of $t$.
The precise normalization of the octet and decuplet baryons are chosen such that in the $SU(3)_f$
limit, the mass sum rules are the same within a multiplet.

Note that, all the currents except the current of $\Lambda$ can be obtained from the currents
of $\Sigma^0$ and $\Sigma^{*0}$ by simple substitutions. Hence, in the following we will give only
the expressions for the $\Sigma^{*0} \rightarrow \Sigma^0$. The expressions for the other decays
can be obtained using the following relationships:
\begin{eqnarray}
\Pi^{\Sigma^{*+} \rightarrow \Sigma^+} &=& \Pi^{\Sigma^{*0} \rightarrow \Sigma^0}(d \rightarrow u)
\nonumber \\
\Pi^{\Sigma^{*-} \rightarrow \Sigma^-} &=& \Pi^{\Sigma^{*0} \rightarrow \Sigma^0}(u \rightarrow d)
\nonumber \\
\Pi^{\Delta^+ \rightarrow p} &=& \Pi^{\Sigma^{*+} \rightarrow \Sigma^+}(s \rightarrow d)
\nonumber \\
\Pi^{\Delta^0 \rightarrow n} &=& \Pi^{\Sigma^{*-} \rightarrow \Sigma^-}(s \rightarrow u)
\nonumber \\
\Pi^{\Xi^{*0} \rightarrow \Xi^0} &=& \Pi^{\Delta^{0} \rightarrow n}(d \rightarrow s)
\nonumber \\
\Pi^{\Xi^{*-} \rightarrow \Xi^-} &=& \Pi^{\Xi^{*0} \rightarrow \Xi^0}(u \rightarrow d)
\label{subs}
\end{eqnarray} 

Recently, it has been shown in \cite{zamir1, zamir2} that it is also possible to obtain the
$\Lambda$ current from that of the $\Sigma^0$ current by substitutions.
For this purpose, note that
\begin{eqnarray} 
2 \eta^{\Sigma^0}(d \leftrightarrow s) = - \sqrt3 \eta^\Lambda - \eta^{\Sigma^0}
\end{eqnarray}
and hence
\begin{eqnarray}
-\sqrt3 \Pi^{\Sigma^{*0} \rightarrow \Lambda} = 
2 \Pi^{\Sigma^{*0} \rightarrow \Sigma^0}(d \leftrightarrow s) + 
\Pi^{\Sigma^{*0} \rightarrow \Sigma^0}
\end{eqnarray}

The correlator function receives three different types of contributions: perturbative contributions, non-perturbative contributions where
photon is emitted from the freely propagating quark, i.e. at short distances, and
non-perturbative contributions where the photon is emitted at long distances. In order to calculate the contribution of the terms
which come from the long distance emission of the photon, the correlation function is expanded
near to the light cone $x^2=0$.The expansion involves matrix elements of
the nonlocal operators between vacuum and the one photon states, which are expressed
in terms of photon wave functions with increasing twist. In other words all long
distance effects are encoded in the matrix elements 
%
of the form  $\langle \gamma(q) \vert \bar q(x_1) \Gamma q(x_2) \vert 0 \rangle$.

For the propagators of the quarks, we have used the following light quark propagator expanded up to
linear order in the quark mass:
\begin{eqnarray}
S_q(x) &=& \frac{i \not\!x}{2\pi^2 x^4} - \frac{m_q}{4 \pi^2 x^2} - \frac{\langle \bar q q \rangle}{12}
\left(1 - i \frac{m_q}{4} \not\!x \right) - \frac{x^2}{192} m_0^2 \langle \bar q q \rangle 
\left( 1 - i \frac{m_q}{6}\not\!x \right) 
\nonumber \\ &&
 - i g_s \int_0^1 du \left[\frac{\not\!x}{16 \pi^2 x^2} G_{\mu \nu} (ux) \sigma_{\mu \nu} - u x^\mu
G_{\mu \nu} (ux) \gamma^\nu \frac{i}{4 \pi^2 x^2} 
\right. \nonumber \\ && \left.
- i \frac{m_q}{32 \pi^2} G_{\mu \nu} \sigma^{\mu
\nu} \left( \ln \left( \frac{-x^2 \Lambda^2}{4} \right) +
2 \gamma_E \right) \right]
\end{eqnarray}
where $\Lambda$ is the energy cut off separating perturbative and non perturbative regimes.

The matrix elements $\langle \gamma(q) \vert \bar q(x_1) \Gamma q(x_2) \vert 0 \rangle$ can be
expanded on the light cone in terms of photon wave functions (in these expansions, we neglect the quark mass corrections)
\cite{ball:wave}:
\begin{eqnarray}
&&\langle \gamma(q) \vert  \bar q(x) \sigma_{\mu \nu} q(0) \vert  0 \rangle  =
-i e_q \qq (\varepsilon_\mu q_\nu - \varepsilon_\nu q_\mu) \int_0^1 du e^{i \bar u qx} 
\left(\chi \varphi_\gamma(u) + \frac{x^2}{16} \mathbb{A}  (u) \right) 
\nonumber \\ && 
-\frac{i}{2(qx)}  e_q \qq \left[x_\nu \left(\varepsilon_\mu - q_\mu \frac{\varepsilon x}{qx}\right) - 
x_\mu \left(\varepsilon_\nu - q_\nu \frac{\varepsilon x}{q x}\right) \right]
\int_0^1 du e^{i \bar u q x} h_\gamma(u)
\nonumber \\
&&\langle \gamma(q) \vert  \bar q(x) \gamma_\mu q(0) \vert 0 \rangle  =
e_q f_{3 \gamma} \left(\varepsilon_\mu - q_\mu \frac{\varepsilon x}{q x} \right) 
\int_0^1 du e^{i \bar u q x} \psi^v(u)
\nonumber \\
&&\langle \gamma(q) \vert \bar q(x) \gamma_\mu \gamma_5 q(0) \vert 0 \rangle  =
- \frac{1}{4} e_q f_{3 \gamma} \epsilon_{\mu \nu \alpha \beta } \varepsilon^\nu q^\alpha x^\beta
\int_0^1 du e^{i \bar u q x} \psi^a(u)
\nonumber \\
&&\langle \gamma(q) | \bar q(x) g_s G_{\mu \nu} (v x) q(0) \vert 0 \rangle =
-i e_q \qq \left(\varepsilon_\mu q_\nu - \varepsilon_\nu q_\mu \right) 
\int {\cal D}\alpha_i e^{i (\alpha_{\bar q} + v \alpha_g) q x} {\cal S}(\alpha_i)
\nonumber \\
&&\langle \gamma(q) | \bar q(x) g_s \tilde G_{\mu \nu} i \gamma_5 (v x) q(0) \vert 0 \rangle =
-i e_q \qq \left(\varepsilon_\mu q_\nu - \varepsilon_\nu q_\mu \right) 
\int {\cal D}\alpha_i e^{i (\alpha_{\bar q} + v \alpha_g) q x} \tilde {\cal S}(\alpha_i)
\nonumber \\
&&\langle \gamma(q) \vert \bar q(x) g_s \tilde G_{\mu \nu}(v x) \gamma_\alpha \gamma_5 q(0) 
\vert 0 \rangle =
e_q f_{3 \gamma} q_\alpha (\varepsilon_\mu q_\nu - \varepsilon_\nu q_\mu) 
\int {\cal D}\alpha_i e^{i (\alpha_{\bar q} + v \alpha_g) q x} {\cal A}(\alpha_i)
\nonumber \\ 
&&\langle \gamma(q) \vert \bar q(x) g_s G_{\mu \nu}(v x) i \gamma_\alpha q(0) 
\vert 0 \rangle =
e_q f_{3 \gamma} q_\alpha (\varepsilon_\mu q_\nu - \varepsilon_\nu q_\mu) 
\int {\cal D}\alpha_i e^{i (\alpha_{\bar q} + v \alpha_g) q x} {\cal V}(\alpha_i)
\nonumber \\ &&
\langle \gamma(q) \vert \bar q(x) \sigma_{\alpha \beta} g_s G_{\mu \nu}(v x) q(0) \vert 0 \rangle  =
e_q \qq \left\{
        \left[\left(\varepsilon_\mu - q_\mu \frac{\varepsilon x}{q x}\right)\left(g_{\alpha \nu} - 
		\frac{1}{qx} (q_\alpha x_\nu + q_\nu x_\alpha)\right) \right. \right. q_\beta 
\nonumber \\ && -
         \left(\varepsilon_\mu - q_\mu \frac{\varepsilon x}{q x}\right)\left(g_{\beta \nu} - 
		\frac{1}{qx} (q_\beta x_\nu + q_\nu x_\beta)\right) q_\alpha 
\nonumber \\ && -
         \left(\varepsilon_\nu - q_\nu \frac{\varepsilon x}{q x}\right)\left(g_{\alpha \mu} - 
		\frac{1}{qx} (q_\alpha x_\mu + q_\mu x_\alpha)\right) q_\beta 
\nonumber \\ &&+
         \left. \left(\varepsilon_\nu - q_\nu \frac{\varepsilon x}{q.x}\right)\left( g_{\beta \mu} - 
		\frac{1}{qx} (q_\beta x_\mu + q_\mu x_\beta)\right) q_\alpha \right]
   \int {\cal D}\alpha_i e^{i (\alpha_{\bar q} + v \alpha_g) qx} {\cal T}_1(\alpha_i) 
\nonumber \\ &&+
        \left[\left(\varepsilon_\alpha - q_\alpha \frac{\varepsilon x}{qx}\right)
		\left(g_{\mu \beta} - \frac{1}{qx}(q_\mu x_\beta + q_\beta x_\mu)\right) \right. q_\nu 
\nonumber \\ &&-
         \left(\varepsilon_\alpha - q_\alpha \frac{\varepsilon x}{qx}\right)
		\left(g_{\nu \beta} - \frac{1}{qx}(q_\nu x_\beta + q_\beta x_\nu)\right)  q_\mu 
\nonumber \\ && -
         \left(\varepsilon_\beta - q_\beta \frac{\varepsilon x}{qx}\right)
		\left(g_{\mu \alpha} - \frac{1}{qx}(q_\mu x_\alpha + q_\alpha x_\mu)\right) q_\nu 
\nonumber \\ &&+
         \left. \left(\varepsilon_\beta - q_\beta \frac{\varepsilon x}{qx}\right)
		\left(g_{\nu \alpha} - \frac{1}{qx}(q_\nu x_\alpha + q_\alpha x_\nu) \right) q_\mu
		\right] 
    \int {\cal D} \alpha_i e^{i (\alpha_{\bar q} + v \alpha_g) qx} {\cal T}_2(\alpha_i)
\nonumber \\ &&+
        \frac{1}{qx} (q_\mu x_\nu - q_\nu x_\mu)
		(\varepsilon_\alpha q_\beta - \varepsilon_\beta q_\alpha)
    \int {\cal D} \alpha_i e^{i (\alpha_{\bar q} + v \alpha_g) qx} {\cal T}_3(\alpha_i) 
\nonumber \\ &&+
        \left. \frac{1}{qx} (q_\alpha x_\beta - q_\beta x_\alpha)
		(\varepsilon_\mu q_\nu - \varepsilon_\nu q_\mu)
    \int {\cal D} \alpha_i e^{i (\alpha_{\bar q} + v \alpha_g) qx} {\cal T}_4(\alpha_i)
                        \right\}
\end{eqnarray}
where $\chi$ is the magnetic susceptibility of the quarks, $\varphi_\gamma(u)$ is the leading
twist 2, $\psi^v(u)$, $\psi^a(u)$, ${\cal A}$ and ${\cal V}$ are the twist 3 and $h_\gamma(u)$,
$\mathbb{A}$, ${\cal T}_i$ ($i=1,~2,~3,~4$) are the twist 4 photon distribution amplitudes.

With all these input, after tedious calculations one obtains the expression of the correlation
function in Eq. (\ref{eq1}) in terms of QCD parameters. The two expressions of the correlation
function, in two different kinematical regions are then matched using dispersion relation. The
contributions of the higher states and the continuum are modeled using the quark-hadron duality.
Finally, the
sum rules are obtained after applying Borel transformations to the results in order both to suppress
the contributions of the higher states and the continuum and to eliminate the polynomials on $p^2$
or $(p+q)^2$ which appear in dispersion relations.

Our final results for the coefficients defined in
Eq. (\ref{sigma}) are:

\begin{eqnarray}
&& -\sqrt3 \lambda_{\Sigma^{*0}} \lambda_{\Sigma^0} (M_{\Sigma^{*0}}+m_{\Sigma^0}) \Sigma_6 =
\nonumber \\ &&
\frac{1}{64 \pi^4}(-1+t) \left( e_u + e_d - 2 e_s \right) M^6
\nonumber \\ &&
- \frac{1}{24 \pi^2}(-1+t) \left( e_u + e_d - 2 e_s \right) f_{3 \gamma} i_1'({\cal A},1-v) M^4
\nonumber \\ &&
- \frac{1}{16 \pi^2} \left[ e_u \uu \left(2 m_d t + m_s (1+t) \right) 
+ e_d \dd \left(2 m_u t + m_s (1+t) \right) 
\right. \nonumber \\ && \left.
- e_s \ss (m_u+m_d) (1+3 t) \right] M^4 \chi \varphi_\gamma( u_0)
\nonumber \\ &&
+ \frac{1}{192 \pi^2} (1 - t) \left( e_u + e_d -  2 e_s \right) f_{3 \gamma} \left[
6 \psi^a( u_0) - \bar u_0 {\psi^a}'( u_0) + 4 \bar u_0 \psi^v( u_0) \right] M^4 
\nonumber \\ &&
- \frac{m_0^2}{108 M^2} \bar u_0 \left[ e_u \uu \left[2 \ss (5 + 2 t) - \dd (5 - t) \right] 
\right. \nonumber \\ && \left.
				+ 	e_d \dd \left[2 \ss (5 + 2 t) - \uu (5 - t) \right] -
					5 e_s \ss (\uu + \dd ) \right] \tilde i_2 (h_\gamma)
\nonumber \\ &&
- \frac{m_0^2}{432 M^2} (1 - t) \bar u_0 \left( e_u (m_d \dd + m_s \ss) + e_d (m_u \uu + m_s \ss) 
\right. \nonumber  \\ && \left.
- 2 e_s (m_u \uu + m_d \dd) \right)
f_{3 \gamma} \left(4 \psi^v(u_0)  - {\psi^a}'(u_0) \right)
\nonumber \\ &&
- \frac{m_0^2}{216 M^2} (1+t) \left[ e_u (\dd m_s + \ss m_d) + e_d (\ss m_u + \uu m_s) 
\right. \nonumber \\ && \left.
- 2 e_s (\dd m_u + \uu m_d) \right] f_{3 \gamma} \psi^a(u_0)
\nonumber \\ &&
+ \frac{1}{8 \pi^2} \left[ e_u m_u \left( \ss (1+t) + 2 \dd t \right) +
			   e_d m_d \left( \ss (1+t) + 2 \uu t \right) 
\right. \nonumber \\ && \left.
		-	   e_s m_s \left(\uu + \dd \right) (1+3 t) \right]
		M^2 \left( \gamma_E - \ln \frac{M^2}{\Lambda^2} \right)
\nonumber \\ && 
- \frac{m_0^2}{144 \pi^2} \left[ e_u \left[ \ss (1+t) (m_d + 5 m_u) + \dd (m_s (1+t) + 10 m_u t) \right] 
				 \right. \nonumber \\ && 
				 +e_d \left[ \ss (1+t) (m_u + 5 m_d) + \uu (m_s (1+t) + 10 m_d t) \right] 
				 \nonumber \\ && \left.
				  - e_s \left[ 2 (1+t) (m_u \dd + m_d \uu) + 5 (1+3 t) m_s (\dd + \uu) \right]
			\right]
		\left( \gamma_E - \ln \frac{M^2}{\Lambda^2} \right)
\nonumber \\ && 
-\frac{m_s}{24 \pi^2} (e_u \uu + e_d \dd) i_1 \left( (1+t) ({\cal T}_1+{\cal T}_3) - (3 + t) ({\cal T}_2 - \tilde{\cal S}) 
- (1 + 3 t) ({\cal T}_4+{\cal S}),1 \right) \times
\nonumber \\ && 
\times M^2 \left( \gamma_E - \ln \frac{M^2}{\Lambda^2} \right)
\nonumber \\ && 
+ e_s \frac{\ss}{24 \pi^2} (m_u + m_d) i_1 \left( (3+t){\cal T}_1 -(1+t) ({\cal T}_2 + {\cal T}_4 + {\cal S} -\tilde {\cal S} ) + (1+ 3t){\cal T}_3,1\right) \times
\nonumber \\ && 
\times M^2 \left( \gamma_E - \ln \frac{M^2}{\Lambda^2} \right)
\nonumber \\ &&
- \frac{1}{12 \pi^2} (e_u \uu m_d + e_d \dd m_u ) i_1 \left( {\cal T}_1 + {\cal T}_2 - \tilde {\cal S} + t ({\cal T}_3 + {\cal T}_4 + {\cal S}),1\right)
M^2 \left( \gamma_E - \ln \frac{M^2}{\Lambda^2} \right)
\nonumber \\ && 
+ e_u \frac{1}{16 \pi^2} \left[ \dd \left(m_d (-1+t) + 4 (m_u - m_s) t \right) + \ss \left(m_s (-1+t) +2 m_u (1+t) - 4 m_d t \right) \right] M^2
\nonumber \\ && 
+ e_d \frac{1}{16 \pi^2} \left[ \uu \left(m_u (-1+t) + 4 (m_d - m_s) t \right) + \ss \left(m_s (-1+t) +2 m_d (1+t) - 4 m_u t \right) \right] M^2
\nonumber \\ && 
- e_s \frac{1}{8 \pi^2} \left[ \uu \left( m_s (1+3t) - m_u (1-t) - 4 m_d t\right) + \dd \left( m_s (1+3 t) -m_d (1-t) -4 m_u t \right) \right] M^2
\nonumber \\ && 
+ \frac{1}{6} \left[ e_u \uu (\ss (1+t) + 2 \dd t) + e_d \dd  (\ss (1+t) + 2 t \uu ) 
\right.\nonumber \\ && \left.
- e_s \ss (1+3 t) (\uu + \dd) \right] M^2 \chi \varphi_\gamma(u_0)
\nonumber \\ &&
- \frac{\bar u_0}{16 \pi^2} \left[e_u \uu (m_s (3+t) - 2 m_d) + e_d \dd (m_s (3+t) -2 m_u ) 
\right. \nonumber \\ && \left.
- e_s \ss (1+t)(m_u+m_d) \right] M^2 \tilde i_2(h_\gamma)
\nonumber \\ &&
+ \frac{1}{32 \pi^2} \left[ e_u \uu (m_s (1+t) + 2 m_d t) + e_d \dd (m_s (1+t) + 2 m_u t) 
\right. \nonumber \\ && \left.
- e_s \ss (m_u + m_d) (1+3 t) \right] M^2 {\mathbb A}( u_0)
\nonumber \\ &&
- e_s \frac{\ss}{24 \pi^2} (m_u+m_d) M^2 \left[ 2 (-1+t) i_1 ({\cal T}_1,1) + 2 (1+t) i_1 ({\cal T}_2,v) + (3 + 5 t) i_1 ({\cal T}_4,1) \right]
\nonumber \\ &&
- \frac{1}{24 \pi^2} \left[ (1+t) m_s (e_d \dd + e_u \uu) - e_s \ss (m_u+m_d) (1+3 t)\right] M^2 \times
\nonumber \\ && \times
\left[ i_1 \left( {\cal T}_2 + 2 {\cal T}_3,1 \right) 
		- 2 i_1 \left({\cal T}_3 - {\cal T}_4,v\right) \right]
\nonumber \\ &&
+\frac{1}{12 \pi^2} \left( e_d \dd m_u + e_u \uu m_d \right) M^2 \left[ - 3 t i_1({\cal S},1) - i_1 (\tilde {\cal S},t - 2 v (1+t)) 
	+ (-1+t) i_1 ({\cal T}_1,1) 
	\right. \nonumber \\ && \left. 
	-i_1({\cal T}_2,t+2 v) - 2 t i_1 ({\cal T}_3,1-v) - t i_1 ({\cal T}_4,1+2 v) \right]
\nonumber \\ &&
+ \frac{m_s}{24 \pi^2} \left( e_d \dd + e_u \uu \right) M^2 \left[ 2 (3+t) i_1 ({\cal T}_2,v) + (3 + 7 t) i_1({\cal T}_4,1) \right]
\nonumber \\ && 
+ \frac{m_s}{24 \pi^2} \left(e_d \dd + e_u \uu \right) M^2 \left[(1+5t) i_1 ({\cal S},1) - (1+t) i_1 (\tilde {\cal S},1) - 4 i_1(\tilde {\cal S},v)
\right]
\nonumber \\ && 
+\frac{t}{3} \left( \ss (e_u \dd + e_d \uu) -2 e_s \uu \dd \right)
\nonumber \\ &&
+ e_s \frac{\ss}{24 \pi^2} (m_u+m_d) M^2 \left[ (-1+t) i_1({\cal S},1) + (1+3 t) i_1(\tilde {\cal S},1) - 4 t i_1 (\tilde {\cal S},v) \right]
\nonumber \\ &&
+ e_u \frac{m_0^2}{144 \pi^2} \left[ \ss (9 m_d t - 2 m_s (-1+t) ) + \dd (9 m_s t - 2 m_d (-1+t) ) \right] 
\nonumber \\ &&
+ e_d \frac{m_0^2}{144 \pi^2} \left[ \ss (9 m_u t - 2 m_s (-1+t) ) + \uu (9 m_s t - 2 m_u (-1+t) ) \right] 
\nonumber \\ &&
- e_s \frac{m_0^2}{72 \pi^2} \left[ \dd (9 m_u t -2 m_d (-1+t)) + \uu (9 m_d t - 2 m_u (-1+t)) \right]
\nonumber \\ &&
- \frac{1}{24} \left[ e_u \uu (\ss (1+t) + 2 \dd t) + e_d \dd (\ss (1+t) + 2 \uu t) 
\right. \nonumber \\ && \left. 
- e_s \ss (1+3t)(\uu+\dd) \right] 
\left({\mathbb A}(u_0) + \frac{10}{9} m_0^2 \chi \varphi_\gamma(u_0) \right)
\nonumber \\ &&
- \frac{\bar u_0}{6} \left( e_u \uu (2 \dd - (3+t)\ss) + e_d \dd (2 \uu - (3+t)\ss) 
\right. \nonumber \\ && \left.
+ e_s \ss (1+t) (\uu + \dd) \right) \tilde i_2(h_\gamma)
\nonumber \\ &&
+ \frac{\bar u_0}{48} (1-t) \left[ e_u (\dd m_d + \ss m_s ) + e_d (\uu m_u + \ss m_s) 
\right. \nonumber \\ && \left. 
-2 e_s (\uu m_u + \dd m_d) \right] 
f_{3 \gamma} \left( 4 \psi^v (\bar u_0) - {\psi^a}'(\bar u_0) \right)
\nonumber \\ &&
+ \frac{e_u}{24} f_{3 \gamma} \left[ \dd (m_d (1-t) +4 m_s t) + \ss (m_s (1-t) + 4 m_d t) \right] \psi^a(u_0)
\nonumber \\ &&
+ \frac{e_d}{24} f_{3 \gamma} \left[ \uu (m_u (1-t) + 4 m_s t) + \ss (m_s (1-t) + 4 m_u t) \right] \psi^a(u_0)
\nonumber \\ &&
- \frac{e_s}{12} f_{3 \gamma} \left[\dd (m_d(1-t) + 4 m_u t) + \uu (m_u (1-t) + 4 m_d t) \right] \psi^a(u_0)
\nonumber \\ &&
+ e_s \frac{\ss}{18} (\uu + \dd) \left[2 (1+t) i_1({\cal T}_2,v) - 2 (1+2t) i_1({\cal T}_2 - {\cal T}_4,1) 
\right. \nonumber \\ && \left. + 
(1+3 t) \left\{i_1({\cal T}_1 - {\cal T}_3,1) + 2 i_1({\cal T}_3 - {\cal T}_4,v) \right\} \right]
\nonumber \\ &&
+ \frac{\ss}{18} (e_u \uu + e_d \dd) \left[ - (1+t) i_1 ({\cal T}_1 - {\cal T}_3,1) 
\right. \nonumber \\ && \left.
-2 (1+t) i_1 ({\cal T}_3 - {\cal T}_4,v)
-2 (1+2t)i_1({\cal T}_4,1) 
\right. \nonumber \\ && \left.
- 2 (3+t) i_1({\cal T}_2,v)  + (2+t) i_1 ({\cal T}_2,1) \right]
\nonumber \\ &&
+ \frac{e_u+e_d}{9} \uu \dd \left[ (-1+t) i_1({\cal T}_2,1) +2 i_1({\cal T}_2,v) - 
t i_1({\cal T}_1 - {\cal T}_3,1) -2 t i_1({\cal T}_3 - {\cal T}_4,v)  \right]
\nonumber \\ &&
+ \frac{t}{9} \left[ e_u \uu (2 \dd - \ss)  + e_d \dd (2 \uu - \ss) - e_s \ss (\uu+\dd) \right] i_1({\cal S},1)
\nonumber \\ &&
+\frac{1}{9} \left[e_u \uu (\dd (1+t) - \ss) + e_d \dd (\uu (1+t) - \ss) 
\right. \nonumber \\ && \left.
-t e_s \ss (\uu+\dd) \right]
i_1(\tilde{\cal S},1-2v)
\nonumber \\ &&
- \frac{f_{3 \gamma}}{36} (1-t) \left[ e_u( \dd m_d - \ss m_s) + e_d (\uu m_u - \ss m_s) \right] i'_1({\cal V},1-v)
\nonumber \\ &&
+ \frac{1}{36}(1-t) \left[e_u (\dd m_d + \ss m_s) + e_d (\uu m_s + \ss m_s) 
\right. \nonumber \\ && \left. 
- 2 e_s (\dd m_d + \uu m_u) \right] f_{3 \gamma} i_1'({\cal A},1-v)
\label{sig6}
\end{eqnarray}

\begin{eqnarray}
&& -\sqrt3 \lambda_{\Sigma^{*0}} \lambda_{\Sigma^0} M_{\Sigma^{*0}} \Sigma_{9} =
- \frac{\bar u_0}{32 \pi^4} (1-t) (e_u + e_d - 2 e_s) M^6
\nonumber \\ &&
- \frac{1}{8 \pi^2} (1+t) \left[ e_u \uu (m_d + m_s) + e_d \dd (m_u + m_s) 
\right. \nonumber \\ && \left.
-2 e_s \ss (m_u + m_d) \right] M^2 \left( M^2 \chi \varphi_\gamma(u_0) - 
\frac12 {\mathbb A}(u_0) \right)
\nonumber \\ &&
+ \frac{1}{24 \pi^2} (1-t) (e_u + e_d - 2 e_s) f_{3 \gamma} \left[ i'_1({\cal A} - {\cal V},1-v)+\frac34 \psi^a(u_0) +3 \tilde i_2(\psi_v) \right] M^4
\nonumber \\ &&
+ \frac{1}{4 \pi^2} (1+t)  \left[ e_u m_u (\dd + \ss) + e_d m_d (\uu + \ss) -  2 e_s m_s (\uu + \dd) \right] \times
 \nonumber \\ &&
\times \left( \gamma_E - \ln \frac{M^2}{\Lambda^2}\right) \left( M^2 - \frac{5}{18} m_0^2 \right)
\nonumber \\ &&
- \frac{m_0^2}{72 \pi^2} (1+t) \left[ e_u (\dd m_s + \ss m_d) + e_d (\uu m_s + \ss m_u)  
\right. \nonumber \\ && \left.
-2 e_s (\dd m_u + \uu m_d) \right]
\left( \gamma_E - \ln \frac{M^2}{\Lambda^2}\right)
\nonumber \\ &&
- \frac{1}{12 \pi^2} (1+t) \left[ e_s \ss (m_u+m_d) i_1\left({\cal S} - \tilde {\cal S} - 2 {\cal T}_1 + {\cal T}_2 - 2 {\cal T}_3 + {\cal T}_4,1 \right)
\right. \nonumber \\ &&
+ (e_u \uu m_d  + e_d \dd m_u) i_1\left({\cal S} - \tilde {\cal S} +{\cal T}_1 + {\cal T}_2 +{\cal T}_3 + {\cal T}_4,1 \right)
\nonumber \\ && \left.
- m_s (e_d \dd + e_u \uu)  i_1\left(2{\cal S} - 2\tilde {\cal S} - {\cal T}_1 + 2{\cal T}_2 - {\cal T}_3 + 2{\cal T}_4,1 \right)
\right] M^2 \left( \gamma_E - \ln \frac{M^2}{\Lambda^2}\right)
\nonumber \\ &&
- \frac{m_0^2}{3 M^2} \bar u_0 \left(e_u \dd \ss + e_d \uu \ss - 2 e_s \uu \dd \right)
\nonumber \\ &&
+ \frac{m_0^2}{216 M^2} (1+t) \left[e_u (\dd m_s + \ss m_d) + e_d (\uu m_s + \ss m_u) 
\right. \nonumber \\ && \left.
-2 e_s (\dd m_u + \uu m_d) \right] 
f_{3 \gamma} \left( 4 \tilde i_2(\psi^v) - \psi^a(u_0) \right)
\nonumber \\ &&
+ \frac{1}{3}(1+t) \left[ e_u \uu (\dd + \ss) +e_d \dd (\uu + \ss) - 2 e_s \ss (\uu + \dd) \right] M^2 \chi \varphi_\gamma(u_0)
\nonumber \\ &&
- \frac{1}{4 \pi^2} \left[ e_u \uu (2 m_d + m_s(1+t)) + e_d \dd (2 m_u + m_s (1+t)) 
\right. \nonumber \\ && \left.
- e_s \ss (m_u + m_d)(3+t) \right]
M^2 \tilde{\tilde i}_2(h_\gamma)
\nonumber \\ &&
+\frac{u_0}{4 \pi^2} (1+t) \left[\dd (2 e_s + e_u)(m_u-m_s) 
+ \uu (2 e_s + e_d)(m_d-m_s)
\right. \nonumber \\ && \left.
+ \ss (e_u-e_d)(m_u-m_d) \right] M^2
\nonumber \\ &&
- \frac{\bar u_0}{4 \pi^2} t \left[m_s (e_u \dd + e_d \uu) -2 e_s (m_u \dd +m_d \uu) + e_s m_s (\uu+\dd)\right] M^2
\nonumber \\ &&
- \frac{\bar u_0}{8 \pi^2} (1-t) \left[ e_u (m_s \ss+m_d \dd)  + e_d (m_s \ss+m_u \uu)- 2 e_s (m_u \uu + m_d \dd) \right] M^2
\nonumber \\ &&
+ \frac{\bar u_0}{4 \pi^2} \left[t e_u \ss (m_u-m_d) + t e_d \ss (m_d-m_u) 
\right. \nonumber \\ && \left.
+  (3+t) (e_u \dd m_u  + e_d \uu m_d - e_s m_s (\uu+\dd) )\right] M^2
\nonumber \\ &&
+ \frac{m_s}{12 \pi^2}(e_u \uu + e_d \dd) \left[ i_1 ({\cal S}- \tilde {\cal S}+ {\cal T}_2+{\cal T}_4,1 + 3 t + 2 v) 
-2 (1+t) i_1 ({\cal T}_3,1-v) \right] M^2
\nonumber \\ &&
+ \frac{1}{12 \pi^2} (e_u \uu m_d + e_d \dd m_u) \left[
(1-3 t) i_1 ({\cal S}+{\cal T}_2,1)
-4 i_1({\cal S}+{\cal T}_2,v)
\right. \nonumber \\ && \left.
+(1+t) i_1 (\tilde {\cal S}-{\cal T}_4,1+2v) 
-2(1+t)i_1({\cal T}_3,1-v) \right] M^2
\nonumber \\ &&
- e_s \frac{\ss}{6 \pi^2} (m_u + m_d) 
 i_1({\cal S}-t \tilde {\cal S}+ {\cal T}_2 - 2 (1+t) {\cal T}_3+ t {\cal T}_4,1-v) M^2
\nonumber \\ &&
+ \frac{1}{3}(1+t) (e_u \dd \ss + e_d \uu \ss - 2 e_s \uu \dd)
\nonumber \\ &&
+ \frac{m_0^2}{16 \pi^2} u_0 (1+t) \left[ e_u (\dd m_s + \ss m_d) + e_d (\uu m_s + \ss m_u) - 2 e_s (\uu m_d + \dd m_u) \right]
\nonumber \\ &&
+ \frac{m_0^2}{72 \pi^2} \bar u_0 (5-t) (e_u m_u + e_d m_d) \ss 
+ \frac{5}{36 \pi^2} e_s m_0^2 m_s \bar u_0(\uu + \dd)
\nonumber \\ &&
- \frac{m_0^2}{72 \pi^2} \bar u_0 (15-t) (e_u \dd m_u + e_d \uu m_d)
\nonumber \\ &&
+ \frac{7 m_0^2}{144 \pi^2} \bar u_0 (1+t)  \left[ e_u (\dd m_s + \ss m_d) + e_d (\uu m_s + \ss m_u) -2 e_s (m_u \dd + m_d \uu) \right]
\nonumber \\ &&
+ \frac{m_0^2}{72 \pi^2} \bar u_0 (1-t) \left[ e_u (\dd m_d + \ss m_s) + e_d (\uu m_u + \ss m_s) - 2 e_s (\uu m_u + \dd m_d) \right]
\nonumber \\ &&
- \frac{1}{12} (1+t) \left[e_u \uu (\dd + \ss ) + e_d \dd (\uu + \ss) -2 e_s \ss (\uu+\dd) \right] \times
 \nonumber \\ && 
\times \left({\mathbb A}(u_0)+ \frac{10}{9}m_0^2 \chi \varphi_\gamma(u_0) \right)
\nonumber \\ &&
+ \frac{1}{6} \left[ e_u (\dd m_s + \ss m_d) + e_d (\uu m_s + \ss m_u) - 2 e_s (\dd m_u + \uu m_d) \right] \times
\nonumber \\ && 
\times f_{3 \gamma}  \left( t \psi^a(u_0) -4 \tilde i_2(\psi^v) \right)
\nonumber \\ &&
+ \frac{1-t}{24} \left[e_u (\dd m_d  + \ss m_s) + e_d (\uu m_u + \ss m_s) - 2 e_s (\dd m_d + \uu m_u) \right] \times
\nonumber \\ &&
\times f_{3 \gamma}  \left( \psi^a(u_0) + 4 \tilde i_2 (\psi^v) \right)
\nonumber \\ &&
+ \frac{1}{3} \left[ e_u \uu (2 \dd + (1+t) \ss) + e_d \dd (2 \uu + (1+t) \ss) 
\right. \nonumber \\ && \left.
- e_s \ss (3+t) (\uu + \dd) \right] \tilde{\tilde i}_2 (h_\gamma)
\nonumber \\ &&
+ \frac{1}{18}(1-t) \left[ e_u \dd m_d + e_d \uu m_u - e_s (\dd m_d + \uu m_u) \right] f_{3 \gamma}i'_1({\cal A-V},1-v)
\nonumber \\ &&
- \frac{1}{9} \ss (e_u \uu + e_d \dd) \left[
i_1({\cal S} - \tilde {\cal S}+{\cal T}_2+{\cal T}_4,-1+t+2v) 
\right. \nonumber \\ && \left.
+ (1+t) i_1 ({\cal T}_1,1)- (1+t) i_1 ({\cal T}_3,1-2v)
\right]
\nonumber \\ &&
-e_s \frac{\ss}{9}(\uu + \dd) \left[
(-1+t) i_1({\cal S}+\tilde {\cal S}+ {\cal T}_2-{\cal T}_4,1)
+2 i_1 ({\cal S}- t \tilde {\cal S}+{\cal T}_2+t{\cal T}_4,v)
\right. \nonumber \\ && \left.
-2 (1+t) i_1 ({\cal T}_1,1)
+2 (1+t) i_1 ({\cal T}_3,1-2v)
\right]
\nonumber \\ &&
+\frac{1}{9} \uu \dd (e_u + e_d) \left[
2 (-1+t) i_1 ({\cal S}+{\cal T}_2,1)
+4 i_1({\cal S}+{\cal T}_2,v)
-2 (1+t) i_1 (\tilde{\cal S}-{\cal T}_4,v)
\right. \nonumber \\ && \left.
-(1+t)i_1 ({\cal T}_1,1)
+(1+t) i_1({\cal T}_3,1-2 v)
\right]
\label{sig9}
\end{eqnarray}

\begin{eqnarray}
&& -\sqrt3 \lambda_{\Sigma^{*0}} \lambda_{\Sigma^0} \Sigma_{12} =
\nonumber \\ &&
-\frac{u_0 \bar u_0}{128\pi^4} (1-t) \left(e_u + e_d - 2 e_s \right) M^4
\nonumber \\ &&
+ \frac{1}{6 \pi^2}(1-t) \left[ (e_d \dd m_u + e_u \uu m_d) \tilde i_1 \left({\cal T}_1 + {\cal T}_2 - {\cal T}_3 - {\cal T}_4,1\right)
\right. \nonumber \\ && \left.
- m_s \left(e_u \uu + e_d \dd \right) \tilde i_1 \left({\cal T}_2 - {\cal T}_4,1\right) 
\right. \nonumber \\ && \left.
- e_s \ss (m_u + m_d) \tilde i_1\left({\cal T}_1 - {\cal T}_3,1 \right) \right] \left( \gamma_E - \ln \frac{M^2}{\Lambda^2} \right)
\nonumber \\ &&
+ \frac{m_0^2}{72 \pi^2 M^2}(1-t) u_0 \bar u_0 \left( e_u (\dd m_d + \ss m_s)  
\right. \nonumber \\ && \left. 
+ e_d (\uu m_u + \ss m_s) -2 e_s (\uu m_u + \dd m_d) \right)
\nonumber \\ &&
+ \frac{\bar u_0}{24 M^2} (1-t) \left( e_u (\dd m_d  + \ss m_s) + e_d (\uu m_u + \ss m_s) 
\right. \nonumber \\ && \left.
-2 e_s (\uu m_u + \dd m_d) \right) f_{3 \gamma} 
\left( \psi^a(u_0) - 4 \tilde i_2(\psi^v) \right)
\nonumber \\ &&
+ \frac{\bar u_0}{3 M^2} \left[ e_u \uu (2 \dd - (3+t)\ss) + e_d \dd (2 \uu - (3+t) \ss) 
\right. \nonumber \\ && \left.
+ e_s \ss (\uu + \dd) (1+t) \right] \tilde {\tilde i}_2(h_\gamma)
\nonumber \\ &&
+\frac{2}{9 M^2} \left[ e_s \ss (1+t) (\dd + \uu) \tilde i_1 \left({\cal T}_2 - {\cal T}_4,1-v \right) 
\right. \nonumber \\ && \left.
+ \uu \dd  (e_u + e_d) \tilde i_1 \left({\cal T}_2 - {\cal T}_4, 1+t-2 v\right) 
\right. \nonumber \\ && \left.
- \ss (e_d \dd + e_u \uu) \tilde i_1 \left( {\cal T}_2 - {\cal T}_4,2 + 2 t - (3+t)v \right)
\right]
\nonumber \\ &&
- \frac{f_{3 \gamma}}{18M^2} (1-t) \left[ (e_u + e_d) m_s \ss \left( i_1({\cal A}-{\cal V},1) + 2 i_1 ({\cal V},1-v) \right)
\right. \nonumber \\ && \left.
- \left(e_u m_d \dd + e_d m_u \uu \right) i_1 ({\cal A}-{\cal V},v)
\right. \nonumber \\ && \left.
- e_s \left( \uu m_u + \dd m_d \right) i_1 ({\cal A}+{\cal V},1-v) \right]
\nonumber \\ &&
+\frac{m_0^2}{216 M^4} (1-t) \bar u_0 \left( e_u (\dd m_d + \ss m_s) + e_d (\uu m_u + \ss m_s) 
\right. \nonumber \\ && \left.
- 2 e_s (\uu m_u + \dd m_d) \right)
f_{3 \gamma} \left( 4 \tilde i_2(\psi^v) - \psi^a( u_0) \right)
\nonumber \\ &&
+ \frac{m_0^2}{54 M^4} \bar u_0 \left[(-5+t) (e_u + e_d) \uu \dd - 5 e_s (1+t) \ss (\uu + \dd) 
\right. \nonumber \\ && \left.
+ 2 \ss (5+2t)(e_d \dd + e_u \uu) \right]
\tilde{\tilde i}_2(h_\gamma)
\nonumber \\ &&
- \frac{1}{96 \pi^2} (1-t)\bar u_0 \left(e_u + e_d - 2 e_s\right) M^2 f_{3 \gamma} \left(  4 \tilde i_2(\psi^v) - \psi^a(\bar u_0) \right)
\nonumber \\ &&
- \frac{1}{24 \pi^2} (1-t)\left( e_u + e_d - 2 e_s\right) M^2 f_{3 \gamma} i_1({\cal A} + {\cal V},1-v)
\nonumber \\ &&
- \frac{1}{6\pi^2} \left[ (e_d \dd m_u + e_u \uu m_d) \left( (-1+t) \tilde i_1({\cal T}_1 - {\cal T}_3,1) + 
2 \tilde i_1({\cal T}_2 - {\cal T}_4,t-v) \right) 
\right. \nonumber \\ && 
- e_s \ss (m_u + d_d) \left( (-1+t) \tilde i_1({\cal T}_1-{\cal T}_3,1) - (1+t) \tilde i_1({\cal T}_2-{\cal T}_4,1-v) \right) 
\nonumber \\ && \left.
- m_s (e_d \dd + e_u \uu) \left( (1+3t) \tilde i_1({\cal T}_2-{\cal T}_4,1) -(3+t) \tilde i_1({\cal T}_2-{\cal T}_4,v) \right) 
\right]
\nonumber \\ &&
- \frac{1}{16 \pi^2} u_0 \bar u_0 (1-t) \left( e_u  (\dd m_d + \ss m_s) + e_d (\uu m_u + \ss m_s) - 2 e_s (\uu m_u + \dd m_d) \right)
\nonumber \\ &&
- \frac{\bar u_0}{8\pi^2} \left[ e_u \uu (2 m_d - (3+t) m_s) + e_d \dd (2 m_u- (3+t) m_s) 
\right. \nonumber \\ && \left.
+ e_s \ss (m_u+m_d) (1+t)\right]
\tilde{\tilde i}_2(h_\gamma)
\nonumber \\ &&
\label{sig12}
\end{eqnarray}
where $\frac{1}{M^2} = \frac{1}{M_1^2} + \frac{1}{M_2^2}$, $M_i^2$ are the Borel parameters.
The contributions of the continuum and the higher states are subtracted using the quark hadron duality by replacing
$M^{2n}$ by $M^{2n} E_n(x)$ and replacing $M^{2n} \ln \frac{M^2}{\Lambda^2}$ by
$M^{2n} \left( \ln \frac{M^2}{\Lambda^2} - E_n(x,M^2) \right) $ for $n > 0$, where $x  = \frac{s_0}{M^2}$, $s_0$ is the continuum threshold,
\begin{eqnarray}
E_n(x) &=& 1 - e^{-x} \sum_{i=0}^n \frac{x^i}{i!}
\nonumber \\ 
E_n(x,M^2) &=& \frac{1}{\Gamma(n)} \int_x^\infty ds s^{n-1} \left( \ln \frac{s M^2}{\Lambda^2} - \psi(n) \right)
\end{eqnarray}
where $\psi(n)$ is the digamma function.

The functions $i_n, ~i_n',~ \tilde i_n, ~\tilde {\tilde i}_n$ appearing in Eqs. (\ref{sig6}-\ref{sig12}) are defined as:
\begin{eqnarray}
i_1(\varphi,f(v)) &=& \int {\cal D}\alpha_i \int_0^1 dv \varphi(\alpha_{\bar q},\alpha_q,\alpha_g) f(v)
\delta(k - \bar u_0) 
\nonumber \\
i_1'(\varphi,f(v)) &=& \int {\cal D}\alpha_i \int_0^1 dv \varphi(\alpha_{\bar q},\alpha_q,\alpha_g)f(v)
\delta'(k - \bar u_0)
\nonumber \\
i_1''(\varphi,f(v)) &=& \int {\cal D}\alpha_i \int_0^1 dv\varphi(\alpha_{\bar q},\alpha_q,\alpha_g)f(v)
\delta''(k - \bar u_0)
\nonumber \\
\tilde i_1(\varphi,f(v)) &=& \int{\cal D}\alpha_i \int_0^1 dv \int_0^{\alpha_{\bar q} + v \alpha_g} dk
\varphi(\alpha_{\bar q},\alpha_q,\alpha_g) f(v) \delta(k-\bar u_0)
\nonumber \\
\tilde i_2(f) &=& \int_{\bar u_0}^1 du f(u)
\nonumber \\
\tilde{\tilde i}_2(f) &=& \int_{\bar u_0}^1 du (u-\bar u_0) f(u)
\end{eqnarray} 
where $k= \alpha_{\bar q} + v \alpha_g$ when there is no integration over $k$, $u_0 = \frac{M_1^2}{M_1^2 + M_2^2}$ and 
$\int {\cal D} \alpha_i = \int_0^1 d \alpha_{\bar q} \int_0^1 d \alpha_q \int_0^1 
d \alpha_g \delta(1-\alpha_{\bar q}-\alpha_q-\alpha_g)$.
Since, for the reactions under considerations, $m \simeq M$, we will set the Borel parameters to be
equal, i.e $M_1^2 =
M_2^2= 2 M^2$ and $u_0 = \frac12$.

In order to obtain the values of the moments $G_X$, ($X=1,~2,~E,~M$), we also need the expression for the
residues. The residues can be obtained from the mass sum rules. For completeness, we also include the mass sum
rules that we used to obtain the values of the residues \cite{hwang, fxlee1}:
\begin{eqnarray}
\lambda_{\Sigma^0}^2 e^{-\frac{m_\Sigma^2}{M^2}} &=&
\frac{M^6}{1024 \pi^2} (5+2t+5t^2)E_2(x) - \frac{m_0^2}{96 M^2}(-1+t)^2 \uu \dd 
\nonumber \\ &&
- \frac{m_0^2}{16 M^2} (-1+t^2) \ss (\dd + \uu)
\nonumber \\ &&
+\frac{3}{128 m_0^2}m_0^2 (-1+t^2) \left[ m_s (\dd + \uu)  + \ss (m_u + m_d) \right]
\nonumber \\ &&
- \frac{1}{64 \pi^2}(-1+ t)^2 \left(\dd m_u + \uu m_d \right) M^2 E_0(x)
\nonumber \\ &&
- \frac{3}{64 \pi^2} (-1+t^2) \left( m_s (\dd + \uu)  + \ss (m_u + m_d ) \right)  M^2 E_0(x)
\nonumber \\ &&
+\frac{1}{128 \pi^2} (5 + 2 t + 5 t^2) \left( \uu m_u + \dd m_d + \ss m_s \right)
\nonumber \\ &&
+ \frac{1}{24} \left[ 3 \ss (\dd + \uu) \left(-1+t^2\right) + (-1+t)^2 \uu \dd  \right]
\nonumber \\ &&
+ \frac{m_0^2}{256 \pi^2} (-1+t)^2 \left(m_d \uu + m_u \dd \right)
\nonumber \\ &&
+ \frac{m_0^2}{256 \pi^2} (-1+t^2) \left[ 13 m_s (\uu + \dd )+ 11 \ss (m_d + m_u) \right]
\nonumber \\ &&
- \frac{m_0^2}{192 \pi^2} (1+t+t^2) \left(\uu m_u +  \dd m_d -2 m_s \ss \right)
\label{masssigma0}
\end{eqnarray}
and
\begin{eqnarray}
M_{\Sigma^{*0}} \lambda_{\Sigma^{*0}}^2 e^{-\frac{m_\Sigma^2}{M^2}} &=& 
\left( \uu + \dd + \ss \right) \frac{M^4}{9 \pi^2} E_1(x) 
- \left( m_u + m_d + m_s\right) \frac{M^6}{32 \pi^4} E_2(x)
\nonumber \\ &&
- \left( \uu + \dd + \ss \right) m_0^2 \frac{M^2}{18 \pi^2} E_0(x) 
\nonumber \\ &&
- \frac{2}{3}\left(1 + \frac{5 m_0^2}{72 M^2} \right) \left( m_u \dd \ss + m_d \ss \uu + m_s \dd \uu \right)
\nonumber \\ &&
+ \left( m_s \dd \ss + m_u \dd \uu + m_d \ss \uu \right) \frac{m_0^2}{12 M^2}
\end{eqnarray}
where $x = \frac{s_0}{M^2}$. (The mass sum rule for the $\Lambda$ baryon can be obtained from Eq. (\ref{masssigma0}) using the relation given in
\cite{zamir2})

Note that, from the mass sum rules, one can only obtain the square of the residues. Hence, from the
mass sum rules it is not possible to deduce the sign of the residues. Thus, in this work, we can not
predict the absolute sign of the moments. For comparison with the experimental data, 
we will also consider the following ratio: 
\begin{eqnarray}
{\cal R}_{EM} = - \frac{G_E}{G_M}
\end{eqnarray}
Although we do not predict the sign of $G_E$ and $G_M$ separately, the sign of ${\cal R}_{EM}$ 
is predicted by light cone QCD sum rules. Our sign convention for the residues are such that $\lambda_{\cal D} > 0$ and $\lambda_{\cal O} > 0$ as $t
\rightarrow \infty$.


An interesting limit to consider is the $SU(3)_f$ limit, i.e. the limit $m_u=m_d=m_s$ and $\uu =
\dd = \ss$, which allows one to establish relationship between the transition amplitudes. 
In this limit,  Eqs. (\ref{sig6}-\ref{sig12}) are proportional to $e_u + e_d - 2 e_s$.
Using Eq. (\ref{subs}), one obtains the following relations:
\begin{eqnarray}
\Sigma^{\Xi^{*-} \rightarrow \Xi^-} &=& \Sigma^{\Sigma^{*-} \rightarrow \Sigma^-} = 0 \nonumber \\
2 \Sigma^{\Sigma^{*0} \rightarrow \Sigma^0} &=& -\sqrt3 \Sigma^{\Sigma^{*0} \rightarrow \Lambda} =
\Sigma^{\Delta^+ \rightarrow p} =- \Sigma^{\Delta^0
\rightarrow n} = \Sigma^{\Sigma^{*+} \rightarrow \Sigma^+} = - \Sigma^{\Xi^{*0} \rightarrow \Xi^0}
\nonumber \\
\label{su3}
\end{eqnarray}
which are the well known $SU(3)_f$ relationships. Note that, Eq. (\ref{su3}) imply that ${\cal R}_{EM}$ is the
same for all considered processes. Here, we should remind once more that the signs are not
predictions of the LCQSR, only relative factors up to a sign are reproduced by the LCQSR. The original
sign convention is chosen such that the $SU(3)$ relations are  reproduced including the signs.
Since we are neglecting the masses of the $u$ and $d$ quarks and the differences in their condensates, isospin subgroup of $SU(3)_f$ remains unbroken.
Hence, the relations $\Sigma^{\Delta^+ \rightarrow p} =- \Sigma^{\Delta^0\rightarrow n}$ still holds in the
approximation that we are considering. 

At the end of this section, we present the decay width for the decay ${\cal D} \rightarrow {\cal O}
\gamma$ in terms of the multipole moments $G_E$ and $G_M$:
\begin{eqnarray}
\Gamma_\gamma = 3 \frac{\alpha}{32} \frac{(M_{\cal D}^2-m_{\cal O}^2)^3}{M_{\cal D}^3 m_{\cal O}^2} \left( G_M^2 + 3 G_E^2\right)
\end{eqnarray}
or in terms of the helicity amplitudes used by PDG \cite{pdg}:
\begin{eqnarray}
\Gamma_\gamma = \frac{w^2}{\pi} \frac{m_{\cal O}}{2 M_{\cal D}} \left( A_{1/2}^2 + A_{3/2}^2 \right)
\end{eqnarray}
where $w$ is the energy of the photon and the helicity amplitudes are defined as:
\begin{eqnarray}
A_{1/2} &=& - \eta \left(G_M - 3 G_E \right) \nonumber \\
A_{3/2} &=& - \sqrt3 \eta \left( G_M + G_E \right)
\end{eqnarray}
where
$$\eta = \frac12 \sqrt\frac{3}{2} \left( \frac{M_{\cal D}^2 - m_{\cal O}^2}{2m_{\cal O}} \right)^{1/2} \frac{e}{2m_{\cal O}}$$

\section{Numerical Analysis}
For the numerical values of the input parameters, the following values are used:
$\uu(1~GeV) = \dd(1~GeV)= -(0.243)^3~GeV^3$, $\ss(1~GeV) = 0.8 \uu(1~GeV)$, $m_0^2(1~GeV) = 0.8$
\cite{belyaev1},
$\chi(1~GeV)=-4.4~GeV^{-2}$ \cite{belyaev2}, $\Lambda = 300~MeV$ and $f_{3 \gamma} = - 0.0039~GeV^2$ \cite{ball:wave}. The photon
wave functions are: \cite{ball:wave}
\begin{eqnarray}
\varphi_\gamma(u) &=& 6 u \bar u \left( 1 + \varphi_2(\mu) C_2^{\frac{3}{2}}(u - \bar u) \right)
\nonumber \\
\psi^v(u) &=& 3 \left(3 (2 u - 1)^2 -1 \right)+\frac{3}{64} \left(15 w^V_\gamma - 5 w^A_\gamma\right) 
                        \left(3 - 30 (2 u - 1)^2 + 35 (2 u -1)^4 \right)
\nonumber \\
\psi^a(u) &=& \left(1- (2 u -1)^2\right)\left(5 (2 u -1)^2 -1\right) \frac{5}{2} 
	\left(1 + \frac{9}{16} w^V_\gamma - \frac{3}{16} w^A_\gamma \right)
\nonumber \\ 
{\cal A}(\alpha_i) &=& 360 \alpha_q \alpha_{\bar q} \alpha_g^2 
		\left(1 + w^A_\gamma \frac{1}{2} (7 \alpha_g - 3)\right)
\nonumber \\
{\cal V}(\alpha_i) &=& 540 w^V_\gamma (\alpha_q - \alpha_{\bar q}) \alpha_q \alpha_{\bar q}
				\alpha_g^2
\nonumber \\
h_\gamma(u) &=& - 10 \left(1 + 2 \kappa^+\right) C_2^{\frac{1}{2}}(u - \bar u) 
\nonumber \\ 
\mathbb{A}(u) &=& 40 u^2 \bar u^2 \left(3 \kappa - \kappa^+ +1\right) 
\nonumber \\ && +
		8 (\zeta_2^+ - 3 \zeta_2) \left[u \bar u (2 + 13 u \bar u) \right. 
\nonumber \\ && + \left.
                2 u^3 (10 -15 u + 6 u^2) \ln(u) + 2 \bar u^3 (10 - 15 \bar u + 6 \bar u^2) 
		\ln(\bar u) \right]
\nonumber \\
{\cal T}_1(\alpha_i) &=& -120 (2 \zeta_2 + \zeta_2^+)(\alpha_{\bar q} - \alpha_q) 
		\alpha_{\bar q} \alpha_q \alpha_g 
\nonumber \\ 
{\cal T}_2(\alpha_i) &=& 30 \alpha_g^2 (\alpha_{\bar q} - \alpha_q) 
	\left((\kappa - \kappa^+) + (\zeta_1 - \zeta_1^+)(1 - 2\alpha_g) + 
	\zeta_2 (3 - 4 \alpha_g)\right)
\nonumber \\
{\cal T}_3(\alpha_i) &=& - 120 (3 \zeta_2 - \zeta_2^+)(\alpha_{\bar q} -\alpha_q) 
		\alpha_{\bar q} \alpha_q \alpha_g 
\nonumber \\
{\cal T}_4(\alpha_i) &=& 30 \alpha_g^2 (\alpha_{\bar q} - \alpha_q) 
	\left((\kappa + \kappa^+) + (\zeta_1 + \zeta_1^+)(1 - 2\alpha_g) + 
	\zeta_2 (3 - 4 \alpha_g)\right)
\end{eqnarray} 
The constants appearing in the wave functions are given as \cite{ball:wave} $\varphi_2(1~GeV) = 0$, $w^V_\gamma = 3.8 \pm 1.8$,
$w^A_\gamma = -2.1 \pm 1.0$, $\kappa = 0.2$, $\kappa^+ = 0$, $\zeta_1 = 0.4$, $\zeta_2 = 0.3$,
$\zeta_1^+ = 0$ and $\zeta_2^+ = 0$

Once the input parameter are determined, the next task is to find the continuum threshold, $s_0$, and a
suitable region of the Borel mass, $M^2$, for each of the processes. To find an upper bound for
the Borel mass parameter, $M^2$, we required the continuum contribution to be less then the
contribution of continuum subtracted sum rules, and requiring that the contribution of the highest power of $\frac{1}{M^2}$ be less than 
$20\%$ of the highest power of $M^2$, gave a lower bound on $M^2$. 
Using these constraints we found that for $\Delta \rightarrow  N \gamma$ transitions,  $s_0 = 3-4~GeV^2$ and $0.9~GeV^2<M^2<1.2~GeV^2$; 
for the $\Sigma^* \rightarrow \Sigma$ and  $\Sigma^{*0} \rightarrow \Lambda$ transitions, $s_0=3-5~GeV^2$ and $0.9~GeV^2< M^2 < 1.2~GeV^2$;
and for the $\Xi^* \rightarrow \Xi \gamma$ transitions, $s_0=3-5~GeV^2$ and $1.1~GeV^2<M^2<1.4~GeV^2$.


In Figs. (\ref{ge.deltap.proton.th}) and (\ref{ge.deltap.proton.9.12.th}), we depict the dependence of
$\left|G_E\right|$ obtained from $\Sigma_6$ and $\Sigma_{12}$, and $\Sigma_9$ and $\Sigma_{12}$, respectively, on $\cos(\theta)$,
where $\theta$ is defined through $t=\tan(\theta)$ for the decays $\Delta \rightarrow N$. Two common features of these graphs are that they become large 
near $\cos(\theta) = \pm 1$, and they go to zero at some finite value of $\cos(\theta)$ These behavior can be understood in the following
way: The correlation function Eq. (\ref{eq1}) is a linear function of $t$.
From the correlation function, one can only obtain the product $\lambda_{\cal O}(t) \lambda_{\cal D} G_E$, and hence this product, obtained from
the correlation function, is also a linear function of $t$. In particular, there is a point $t=t_0$ at which this product goes to zero. On the other
hand, from the definition of $\lambda_{\cal O}(t)$, Eq. (\ref{eq4}), it is also seen that the residue is also 
a linear function of $t$, which has to be determined using the mass
sum rules Eq. (\ref{masssigma0}), and consequently there is a point $t=t'_0$ at which the residue goes to zero. If one could make an exact calculation
of the correlation function Eq. (\ref{eq1}) and the mass sum rule Eq. (\ref{masssigma0}), then one should obtain $t_0=t'_0$, but due to
the approximations used, these two points do not coincide. This is reflected in Figs. (\ref{ge.deltap.proton.th}) and (\ref{ge.deltap.proton.9.12.th})
as a point at which $\left| G_E \right|$ go to zero (at $t=t_0$) and a point at which $\left| G_E \right|$ becomes very large (near $t=t'_0$). These points
and any region between them, as well as any enhancement/suppression near these points, is an artifact of the approximations used, and hence the suitable
region for $t$ should be away from these points.

Another point to observe comparing these two figures is that in Fig. (\ref{ge.deltap.proton.th}), the points at which $\left| G_E \right|$ vanishes and
that it gets very large are very close together whereas in Fig. (\ref{ge.deltap.proton.9.12.th}), they are farther apart. In particular, in Fig.
(\ref{ge.deltap.proton.9.12.th}), the Ioffe current, which corresponds to $\cos(\theta) = - \frac1{\sqrt2} \simeq -0.71$, is out of the suitable region
for $t$. For the working region of $\cos(\theta)$, if one chooses $-0.5 \le \cos(\theta) \le 0.5$ for  Fig. (\ref{ge.deltap.proton.th}), and
$-0.2 \le \cos(\theta) \le 0.3$ for  Fig. (\ref{ge.deltap.proton.9.12.th}), one obtains $\left| G_E \right| = 0.17\pm 0.04$ and
$\left| G_E \right| = 0.21 \pm 0.07$ respectively, i.e. within the respective working regions of $t$, using the sum rules for $\Sigma_6$ and $\Sigma_{12}$ or $\Sigma_9$
and $\Sigma_{12}$, the predictions obtained for $G_E$ are in agreement. But due to the better stability of the results obtained using
the sum rules for $\Sigma_6$ and $\Sigma_{12}$ with respect to variations of $t$ and hence a larger working region of $\cos(\theta)$, from now on
we will present only the results of the sum rules obtained from $\Sigma_6$ and $\Sigma_{12}$. 

Figs. (\ref{gm.deltap.proton.th}) and (\ref{gm.deltap.proton.9.12.th}), are the same as Figs. (\ref{ge.deltap.proton.th}) and
(\ref{ge.deltap.proton.9.12.th}), but for $\left| G_M \right|$. The general features of the dependence of $\left| G_M \right|$ on  $\cos(\theta)$ is the
same as those of $\left| G_E \right|$, i.e. Fig. (\ref{gm.deltap.proton.th}) is more stable than Fig. (\ref{gm.deltap.proton.9.12.th}) as a function of
$t$ and hence has a larger working region for $\cos(\theta)$.

In Figs. (\ref{ge.deltap.proton.Msq})-(\ref{gm.xism.xim.Msq}), the dependence of $G_E$
and $G_M$ on $M^2$ for various processes and different values of the continuum threshold is depicted for the Ioffe current and the limit $|t| \rightarrow
\infty$ ($\cos(\theta)=0$). The Ioffe current always gives smaller results due to its vicinity to the point where the correlation function vanishes
as a function of $t$.
From the figures it is seen that all of the sum rules are stable with
respect to small variation of $s_0$ in the region of $M^2$ considered.

In Table (\ref{result}), we show our results for the moments $G_X$
($X=~E,~M$) and also ${\cal R}_{EM}$ values for the central values of $G_E$ and $G_M$. For the values in the table, the value of $\cos(\theta)$ is
restricted to be $-0.5 \le \cos(\theta) \le 0.5$
\begin{table}
\begin{center}
\begin{tabular}{|l|c|c|c|c|c|}
\hline
Process			  	  &      $G_E$ 		& $G_M$ 		& ${\cal R}_{EM}(\%)$ \\
\hline
$\Delta^+ \rightarrow p$  	  &  $0.17\pm0.05$ 	&  $2.5\pm1.3$ 		& $   -6.8   $   \\
\hline
$\Delta^0 \rightarrow n$  	  & $-0.17\pm0.05$ 	& $-2.5\pm1.3$ 		& $   -6.8   $   \\
\hline
$\Sigma^{*+} \rightarrow \Sigma^+$& $0.08\pm0.02$ 	&  $2.1\pm0.85$ 	& $   -3.8   $   \\
\hline
$\Sigma^{*0} \rightarrow \Sigma^0$ &$0.034\pm0.007$   	&$0.89\pm0.38$		& $   -3.8   $   \\
\hline
$\Sigma^{*0} \rightarrow \Lambda$ &  $-0.13\pm0.02$ 	&$-2.3\pm1.4$ 		& $   -5.7   $   \\
\hline
$\Sigma^{*-} \rightarrow \Sigma^-$& $-0.010\pm0.004$ 	&$-0.31\pm0.10$		& $   -3.2   $   \\
\hline
$\Xi^{*0} \rightarrow \Xi^0$	  &  $-0.09\pm0.02$ 	& $-2.2\pm0.74$ 	& $   -4.1   $   \\
\hline
$\Xi^{*-} \rightarrow \Xi^-$	  &$0.011\pm0.003$ 	& $0.31\pm0.11$ 	& $   -3.5   $   \\
\hline
\end{tabular}
\end{center}
\caption{The predictions on the moments for various decays. The magnetic moments are given in
terms of natural magnetons}
\label{result}
\end{table}
The errors quoted are due to the variations of the Borel mass $M^2$, the continuum threshold $s_0$ and the $t$ parameter. The largest uncertainty is due
to the residual dependence on the value of $t$.
The non zero values for the moments of the transitions $\Sigma^{*-} \rightarrow \Sigma^-$ and $\Xi^{*-}
\rightarrow \Xi^-$ are purely due to $SU(3)_f$ violating effects.
For the other decays, within theoretical errors, the values of $G_M$ respect the $SU(3)_f$ flavor symmetry. 

In Table (\ref{comparison}), we present our results and the results obtained from lattice
calculation \cite{leinweber}. Note that, the conventions that we use are different from the
conventions used in \cite{leinweber}. This difference leads to a factor of $\sqrt{3/2}$ between our
results and the results in \cite{leinweber}. Hence, to make a comparison, the values we quote in
Table (\ref{comparison}) are the results given in \cite{leinweber} multiplied by $\sqrt{2/3}$.
We see that for the values of $G_M$, our results are in agreement with the results from
lattice within error bars. The main reason for our large error bars are due to the residual $t$ dependence of
our results, where as in \cite{leinweber}, only the limit $t \rightarrow \infty$ is considered.

For the values of $G_E$, out results are in agreement with the results of lattice calculations for the channels $\Sigma^{*+(0)} \rightarrow \Sigma^{+(0)}$.
and the biggest discrepancy between out results and the lattice results are for the channels $\Sigma^{*-} \rightarrow \Sigma^-$ and $\Xi^{*-} \rightarrow
\Xi^-$, where even within error bars, we do not agree even on the sign of $G_E$. Note that these two channels are the channels for which
in the case of exact $SU(3)_f$ symmetry, $G_E=0$. For the remaining channels $\Delta \rightarrow N$ and $\Xi^{*0} \rightarrow \Xi^0$, there is not
agreements with the results of the lattice calculation but within error, we do agree on the sign of $G_E$ for these channels.
\begin{table}
\begin{tabular}{|l|c|c|c|c|c|c|}
\hline
Process		& $G_E$	& $G_E^{\cite{leinweber}}$ & $G_M$ & $G_M^{\cite{leinweber}}$ & 
 ${\cal R}_{EM}(\%)$ & ${\cal R}_{EM}^{\cite{leinweber}}(\%)$ \\
\hline
$\Delta^+ \rightarrow p$ 	   & $0.17\pm0.05$ &-0.04(11)&$2.5\pm1.3$& 2.01(33) &$-6.8$ & 3(8) \\
\hline
$\Delta^0 \rightarrow n$ 	   &$-0.17\pm0.05$ &0.04(11) &$-2.5\pm1.3$&-2.01(33)&$-6.8$ & 3(8) \\
\hline
$\Sigma^{*+} \rightarrow \Sigma^+$ &$-0.08\pm0.02$  &-0.06(8) &$2.1\pm0.85$& 2.13(16) &$-3.8$& 5(6)\\
\hline
$\Sigma^{*0} \rightarrow \Sigma^0$ &$-0.034\pm0.007$ &-0.02(4)  &$0.89\pm0.38$&0.87(7)&$-3.8$ & 4(6)\\
\hline
$\Sigma^{*-} \rightarrow \Sigma^-$ &$-0.010\pm0.004$ & 0.020(10)&$-0.31\pm0.10$ &-0.38(4) & $-3.2$ & 8(4)\\
\hline
$\Xi^{*0} \rightarrow \Xi^0$ 	   &$-0.09\pm0.02$ & 0.03(4)&$-2.2\pm0.74$&-2.26(14)&$-4.1$ & 2.4(27)\\
\hline
$\Xi^{*-} \rightarrow \Xi^-$ 	   &$0.011\pm0.003$ &-0.018(7) & $0.31\pm0.11$  & 0.38(3) & $-3.5$& 7.4(30)\\
\hline
\end{tabular}
\caption{Our results together with the results from lattice \cite{leinweber}}
\label{comparison}
\end{table}

In Table (\ref{amplitudes}), we give our predictions for the helicity amplitudes and the decay widths
for the corresponding decay. For the time being, the experimental data is available only for the
decays \cite{pdg} $\Delta \rightarrow N \gamma$ for which $A_{1/2} = -0.135\pm0.005~GeV^{-1/2}$, $A_{3/2}
=-0.250\pm0.008~GeV^{-1/2}$ and $\Gamma =0.64\pm0.06~MeV$. 
It is seen that our predictions for the
amplitudes for these decays are in agreement with the experimental results. Recently, SELEX Collaboration has announced an upper bound for the radiative width of the decay $\Sigma^{*-} \rightarrow \Sigma^- \gamma$ as $\Gamma(\Sigma^{*-} \rightarrow \Sigma^- \gamma) < 9.5 ~KeV$\cite{selex}. 
Our prediction for the width of this channel is $\Gamma = 2 \pm1~KeV$ which is below the experimental bound.
\begin{table}
\begin{tabular}{|l|c|c|c|c|c|c|}
\hline
Process	&  				$A_{1/2}(GeV^{-1/2})$	& $A_{3/2}(GeV^{-1/2})$  & $\Gamma(MeV)$ \\ 
\hline
$\Delta^+ \rightarrow p$ 	   	&$-0.12\pm0.09$ &$-0.27\pm0.14$ &$0.90\pm0.73$  \\
\hline
$\Delta^0 \rightarrow n$ 	   	&$0.12\pm0.09$  &$0.27\pm0.14$  &$0.90\pm0.73$  \\
\hline
$\Sigma^{*+} \rightarrow \Sigma^+$ 	&$-0.067\pm0.033$ &$-0.14\pm0.05$ &$0.11\pm0.82$\\
\hline
$\Sigma^{*0} \rightarrow \Sigma^0$ 	&$-0.029\pm0.014$ &$-0.057\pm0.024$ &$0.021\pm0.015$\\
\hline
$\Sigma^{*0} \rightarrow \Lambda$ 	&$0.088\pm0.067$   &$0.20\pm0.11$ &$0.47\pm0.41$  \\
\hline
$\Sigma^{*-} \rightarrow \Sigma^-$ 	&$0.010\pm0.004$ &$0.020\pm0.006$ &$0.002\pm0.001$ \\
\hline
$\Xi^{*0} \rightarrow \Xi^0$ 	   	&$0.067\pm0.028$  &$0.14\pm0.05$  &$0.14\pm0.09$ \\
\hline
$\Xi^{*-} \rightarrow \Xi^-$ 	   	&$-0.010\pm0.004$ &$-0.019\pm0.007$ &$0.003\pm0.002$ \\
\hline
\end{tabular}
\caption{The predictions for the helicity amplitudes and the decay widths for various decays}
\label{amplitudes}
\end{table}

\section*{Acknowledgments}
We would like to thank V.S. Zamiralov for useful discussions on the relationship between the
expressions for the $\Sigma^0$ and $\Lambda$ baryon properties. T.M. would like to thank the High
Energy Section of The Abdus Salam ICTP, Trieste, Italy, and A.O. would like to thank the physics
department of Middle East Technical University, Ankara, Turkey where parts of this work is carried
out for their hospitality.

\newpage
\section*{Figure Captions}

\begin{itemize}
\item[Fig. \ref{ge.deltap.proton.th}] The dependence of $G_E^{\Delta \rightarrow N}$, obtained from $\Sigma_6$ and $\Sigma_{12}$, 
on $\cos(\theta)$ for $s_0=3~GeV^2$ and $s_0=4~GeV^2$ for the value of the Borel parameter $M^2=1~GeV^2$
\item[Fig. \ref{ge.deltap.proton.9.12.th}] 
The same as Fig. \ref{ge.deltap.proton.th} but for $G_E^{\Delta \rightarrow N}$ obtained from $\Sigma_9$ and $\Sigma_{12}$
\item[Fig. \ref{gm.deltap.proton.th}] The same as Fig. \ref{ge.deltap.proton.th} but for $G_M^{\Delta \rightarrow N}$
\item[Fig. \ref{gm.deltap.proton.9.12.th}] The same as Fig. \ref{ge.deltap.proton.9.12.th} but for $G_M^{\Delta \rightarrow N}$
\item[Fig. \ref{ge.deltap.proton.Msq}] The dependence of $G_E^{\Delta^+ \rightarrow p}$, obtained from $\Sigma_6$ and $\Sigma_{12}$,
on the borel parameter $M^2$ for $s0=3~GeV^2$ and $s_0=4~GeV^2$. The lines with circles on them correspond to the Ioffe current, $t=-1$, whereas
others correspond to $t=\infty$
\item[Fig. \ref{gm.deltap.proton.Msq}] The same as Fig. \ref{ge.deltap.proton.Msq} but for $G_M^{\Delta^+ \rightarrow p}$
\item[Fig. \ref{ge.sigmasp.sigmap.th}]The same as Fig. \ref{ge.deltap.proton.th}, but for the decay $\Sigma^{*+}
\rightarrow \Sigma^+$ and for $s0=3,~4,~5~GeV^2$
\item[Fig. \ref{ge.sigmasp.sigmap.Msq}]The same as Fig. \ref{ge.deltap.proton.Msq}, but for the decay $\Sigma^{*+} \rightarrow
\Sigma^+$ and for $s0=3,~4,~5~GeV^2$ 
\item[Fig. \ref{gm.sigmasp.sigmap.th}] The same as Fig. \ref{ge.sigmasp.sigmap.th}, but for $G_M^{\Sigma^{*+}
\rightarrow \Sigma^+}$
\item[Fig. \ref{gm.sigmasp.sigmap.Msq}]The same as Fig. \ref{ge.sigmasp.sigmap.Msq}, but for $G_M^{\Sigma^{*+} \rightarrow
\Sigma^+}$ 
\item[Fig. \ref{ge.sigmas0.sigma0.th}] The same as Fig. \ref{ge.sigmasp.sigmap.th}, but for the decay $\Sigma^{*0}
\rightarrow \Sigma^0$
\item[Fig. \ref{ge.sigmas0.sigma0.Msq}]The same as Fig. \ref{ge.sigmasp.sigmap.Msq}, but for the decay $\Sigma^{*0} \rightarrow
\Sigma^0$
\item[Fig. \ref{gm.sigmas0.sigma0.th}]The same as Fig. \ref{gm.sigmasp.sigmap.th}, but for the decay $\Sigma^{*0}
\rightarrow \Sigma^0$

\item[Fig. \ref{gm.sigmas0.sigma0.Msq}]The same as Fig. \ref{gm.sigmasp.sigmap.Msq}, but for the decay $\Sigma^{*0} \rightarrow
\Sigma^0$

\item[Fig. \ref{ge.sigmas0.lambda.th}]The same as Fig. \ref{ge.sigmasp.sigmap.th}, but for the decay $\Sigma^{*0}
\rightarrow \Lambda$

\item[Fig. \ref{ge.sigmas0.lambda.Msq}]The same as Fig. \ref{ge.sigmasp.sigmap.Msq}, but for the decay $\Sigma^{*0} \rightarrow
\Lambda$

\item[Fig. \ref{gm.sigmas0.lambda.th}]The same as Fig. \ref{gm.sigmasp.sigmap.th}, but for the decay $\Sigma^{*0}
\rightarrow \Lambda$

\item[Fig. \ref{gm.sigmas0.lambda.Msq}]The same as Fig. \ref{gm.sigmasp.sigmap.Msq}, but for the decay $\Sigma^{*0} \rightarrow
\Lambda$

\item[Fig. \ref{ge.sigmasm.sigmam.th}] The same as Fig. \ref{ge.sigmasp.sigmap.th}, but for the decay $\Sigma^{*-}
\rightarrow \Sigma^-$

\item[Fig. \ref{ge.sigmasm.sigmam.Msq}] The same as Fig. \ref{ge.sigmasp.sigmap.Msq}, but for the decay $\Sigma^{*-} \rightarrow
\Sigma^-$

\item[Fig. \ref{gm.sigmasm.sigmam.th}] The same as Fig. \ref{gm.sigmasp.sigmap.th}, but for the decay $\Sigma^{*-}
\rightarrow \Sigma^-$

\item[Fig. \ref{gm.sigmasm.sigmam.Msq}] The same as Fig. \ref{gm.sigmasp.sigmap.Msq}, but for the decay $\Sigma^{*-} \rightarrow
\Sigma^-$

\item[Fig. \ref{ge.xis0.xi0.th}] The same as Fig. \ref{ge.sigmasp.sigmap.th}, but for the decay $\Xi^{*0}
\rightarrow \Xi^0$ and for the Borel parameter $M^2=1.2~GeV^2$

\item[Fig. \ref{ge.xis0.xi0.Msq}] The same as Fig. \ref{ge.sigmasp.sigmap.Msq}, but for the decay $\Xi^{*0} \rightarrow
\Xi^0$ and for the Borel parameter $M^2=1.2~GeV^2$

\item[Fig. \ref{gm.xis0.xi0.th}] The same as Fig. \ref{gm.sigmasp.sigmap.th}, but for the decay $\Xi^{*0}
\rightarrow \Xi^0$ and for the Borel parameter $M^2=1.2~GeV^2$

\item[Fig. \ref{gm.xis0.xi0.Msq}] The same as Fig. \ref{gm.sigmasp.sigmap.Msq}, but for the decay $\Xi^{*0} \rightarrow
\Xi^0$ and for the Borel parameter $M^2=1.2~GeV^2$

\item[Fig. \ref{ge.xism.xim.th}] The same as Fig. \ref{ge.sigmasp.sigmap.th}, but for the decay $\Xi^{*-}
\rightarrow \Xi^-$ and for the Borel parameter $M^2=1.2~GeV^2$

\item[Fig. \ref{ge.xism.xim.Msq}] The same as Fig. \ref{ge.sigmasp.sigmap.Msq}, but for the decay $\Xi^{*-} \rightarrow
\Xi^-$ and for the Borel parameter $M^2=1.2~GeV^2$

\item[Fig. \ref{gm.xism.xim.th}] The same as Fig. \ref{gm.sigmasp.sigmap.th}, but for the decay $\Xi^{*-}
\rightarrow \Xi^-$ and for the Borel parameter $M^2=1.2~GeV^2$

\item[Fig. \ref{gm.xism.xim.Msq}] The same as Fig. \ref{gm.sigmasp.sigmap.Msq}, but for the decay $\Xi^{*-} \rightarrow
\Xi^-$ and for the Borel parameter $M^2=1.2~GeV^2$
\end{itemize}
\newpage
\figt{GE.deltap.proton.th.Msq_1.eps}{ge.deltap.proton.th}{The dependence of $G_E^{\Delta \rightarrow N}$, obtained from $\Sigma_6$ and $\Sigma_{12}$, 
on $\cos(\theta)$ for $s_0=3~GeV^2$ and $s_0=4~GeV^2$ for the value of the Borel parameter $M^2=1~GeV^2$}
\figb{GE.deltap.proton.9.12.th.Msq_1.eps}{ge.deltap.proton.9.12.th}{The same as Fig. \ref{ge.deltap.proton.th} but for $G_E^{\Delta \rightarrow N}$
obtained from $\Sigma_9$ and $\Sigma_{12}$}

\newpage
\figt{GM.deltap.proton.th.Msq_1.eps}{gm.deltap.proton.th}{The same as Fig. \ref{ge.deltap.proton.th} but for $G_M^{\Delta \rightarrow N}$}
\figb{GM.deltap.proton.9.12.th.Msq_1.eps}{gm.deltap.proton.9.12.th}{The same as Fig. \ref{ge.deltap.proton.9.12.th} but for $G_M^{\Delta \rightarrow N}$}

\newpage
\figt{GE.deltap.proton.Msq.eps}{ge.deltap.proton.Msq}{The dependence of $G_E^{\Delta^+ \rightarrow p}$, obtained from $\Sigma_6$ and $\Sigma_{12}$,
on the borel parameter $M^2$ for $s0=3~GeV^2$ and $s_0=4~GeV^2$. The lines with circles on them correspond to the Ioffe current, $t=-1$, whereas 
others correspond to $t=\infty$}
\figb{GM.deltap.proton.Msq.eps}{gm.deltap.proton.Msq}{The same as Fig. \ref{ge.deltap.proton.Msq} but for $G_M^{\Delta^+ \rightarrow p}$}

\newpage
\figt{GE.sigmasp.sigmap.th.Msq_1.eps}{ge.sigmasp.sigmap.th}{The same as Fig. \ref{ge.deltap.proton.th}, but for the decay $\Sigma^{*+}
\rightarrow \Sigma^+$ and for $s0=3,~4,~5~GeV^2$}
\figb{GE.sigmasp.sigmap.Msq.eps}{ge.sigmasp.sigmap.Msq}{The same as Fig. \ref{ge.deltap.proton.Msq}, but for the decay $\Sigma^{*+} \rightarrow
\Sigma^+$ and for $s0=3,~4,~5~GeV^2$ }

\newpage
\figt{GM.sigmasp.sigmap.th.Msq_1.eps}{gm.sigmasp.sigmap.th}{The same as Fig. \ref{ge.sigmasp.sigmap.th}, but for $G_M^{\Sigma^{*+}
\rightarrow \Sigma^+}$}
\figb{GM.sigmasp.sigmap.Msq.eps}{gm.sigmasp.sigmap.Msq}{The same as Fig. \ref{ge.sigmasp.sigmap.Msq}, but for $G_M^{\Sigma^{*+} \rightarrow
\Sigma^+}$}

\newpage
\figt{GE.sigmas0.sigma0.th.Msq_1.eps}{ge.sigmas0.sigma0.th}{The same as Fig. \ref{ge.sigmasp.sigmap.th}, but for the decay $\Sigma^{*0}
\rightarrow \Sigma^0$}
\figb{GE.sigmas0.sigma0.Msq.eps}{ge.sigmas0.sigma0.Msq}{The same as Fig. \ref{ge.sigmasp.sigmap.Msq}, but for the decay $\Sigma^{*0} \rightarrow
\Sigma^0$}

\newpage
\figt{GM.sigmas0.sigma0.th.Msq_1.eps}{gm.sigmas0.sigma0.th}{The same as Fig. \ref{gm.sigmasp.sigmap.th}, but for the decay $\Sigma^{*0}
\rightarrow \Sigma^0$}
\figb{GM.sigmas0.sigma0.Msq.eps}{gm.sigmas0.sigma0.Msq}{The same as Fig. \ref{gm.sigmasp.sigmap.Msq}, but for the decay $\Sigma^{*0} \rightarrow
\Sigma^0$}

\newpage
\figt{GE.sigmas0.lambda.th.Msq_1.eps}{ge.sigmas0.lambda.th}{The same as Fig. \ref{ge.sigmasp.sigmap.th}, but for the decay $\Sigma^{*0}
\rightarrow \Lambda$}
\figb{GE.sigmas0.lambda.Msq.eps}{ge.sigmas0.lambda.Msq}{The same as Fig. \ref{ge.sigmasp.sigmap.Msq}, but for the decay $\Sigma^{*0} \rightarrow
\Lambda$}

\newpage
\figt{GM.sigmas0.lambda.th.Msq_1.eps}{gm.sigmas0.lambda.th}{The same as Fig. \ref{gm.sigmasp.sigmap.th}, but for the decay $\Sigma^{*0}
\rightarrow \Lambda$}
\figb{GM.sigmas0.lambda.Msq.eps}{gm.sigmas0.lambda.Msq}{The same as Fig. \ref{gm.sigmasp.sigmap.Msq}, but for the decay $\Sigma^{*0} \rightarrow
\Lambda$}

\newpage
\clearpage
\figt{GE.sigmasm.sigmam.th.Msq_1.eps}{ge.sigmasm.sigmam.th}{The same as Fig. \ref{ge.sigmasp.sigmap.th}, but for the decay $\Sigma^{*-}
\rightarrow \Sigma^-$}
\figb{GE.sigmasm.sigmam.Msq.eps}{ge.sigmasm.sigmam.Msq}{The same as Fig. \ref{ge.sigmasp.sigmap.Msq}, but for the decay $\Sigma^{*-} \rightarrow
\Sigma^-$}

\newpage
\figt{GM.sigmasm.sigmam.th.Msq_1.eps}{gm.sigmasm.sigmam.th}{The same as Fig. \ref{gm.sigmasp.sigmap.th}, but for the decay $\Sigma^{*-}
\rightarrow \Sigma^-$}
\figb{GM.sigmasm.sigmam.Msq.eps}{gm.sigmasm.sigmam.Msq}{The same as Fig. \ref{gm.sigmasp.sigmap.Msq}, but for the decay $\Sigma^{*-} \rightarrow
\Sigma^-$}

\newpage
\figt{GE.xis0.xi0.th.Msq_1.2.eps}{ge.xis0.xi0.th}{The same as Fig. \ref{ge.sigmasp.sigmap.th}, but for the decay $\Xi^{*0}
\rightarrow \Xi^0$ and for the Borel parameter $M^2=1.2~GeV^2$}
\figb{GE.xis0.xi0.Msq.eps}{ge.xis0.xi0.Msq}{The same as Fig. \ref{ge.sigmasp.sigmap.Msq}, but for the decay $\Xi^{*0} \rightarrow
\Xi^0$ and for the Borel parameter $M^2=1.2~GeV^2$}

\newpage
\figt{GM.xis0.xi0.th.Msq_1.2.eps}{gm.xis0.xi0.th}{The same as Fig. \ref{gm.sigmasp.sigmap.th}, but for the decay $\Xi^{*0}
\rightarrow \Xi^0$ and for the Borel parameter $M^2=1.2~GeV^2$}
\figb{GM.xis0.xi0.Msq.eps}{gm.xis0.xi0.Msq}{The same as Fig. \ref{gm.sigmasp.sigmap.Msq}, but for the decay $\Xi^{*0} \rightarrow
\Xi^0$ and for the Borel parameter $M^2=1.2~GeV^2$}

\newpage
\figt{GE.xism.xim.th.Msq_1.2.eps}{ge.xism.xim.th}{The same as Fig. \ref{ge.sigmasp.sigmap.th}, but for the decay $\Xi^{*-}
\rightarrow \Xi^-$ and for the Borel parameter $M^2=1.2~GeV^2$}
\figb{GE.xism.xim.Msq.eps}{ge.xism.xim.Msq}{The same as Fig. \ref{ge.sigmasp.sigmap.Msq}, but for the decay $\Xi^{*-} \rightarrow
\Xi^-$ and for the Borel parameter $M^2=1.2~GeV^2$}

\newpage
\figt{GM.xism.xim.th.Msq_1.2.eps}{gm.xism.xim.th}{The same as Fig. \ref{gm.sigmasp.sigmap.th}, but for the decay $\Xi^{*-}
\rightarrow \Xi^-$ and for the Borel parameter $M^2=1.2~GeV^2$}
\figb{GM.xism.xim.Msq.eps}{gm.xism.xim.Msq}{The same as Fig. \ref{gm.sigmasp.sigmap.Msq}, but for the decay $\Xi^{*-} \rightarrow
\Xi^-$ and for the Borel parameter $M^2=1.2~GeV^2$}


\newpage

\begin{thebibliography}{99}

\bibitem{R1} 
C. Mertz {\it et al.}, Phys. Rev. Lett. {\bf 86} (2001) 2963

\bibitem{R2}
K. Joo {\it et al.} CLAS Collab. Phys. Rev. Lett. {\bf 88} (2002) 122001

\bibitem{R3}
M. A. Shifman, A.I. Vainshtein, and V. I. Zakharov, Nucl. Phys. {\bf B147} (1979)
385

\bibitem{R4}
P. Colangelo and A. Khodjamirian in ``At the Frontier of Particle Physics: Handbook
of QCD,'' ed. by M. Shifman (World Scientific, Singapore, 2001) V.3, 1995

\bibitem{R5}
V.M. Braun, prep. hep-ph/9801222 (1998)

\bibitem{R6}
M.N. Butler, M.J. Savage and R.P. Springer, Nucl. Phys. {\bf B399} (1993) 69

\bibitem{R7}
M.N. Butler, M.J. Savage and R.P. Springer, Phys. Lett. {\bf B304} (1993) 353

\bibitem{R8}
M. Napsuciale and J. L. Lucio, Nucl. Phys. {\bf B494} (1997) 260

\bibitem{R9}
G.C. Gellas, T.R. Hemmart, C.N. Ktorides and G. I. Poulis, Phys. Rev. {\bf D60}
(1999) 054022

\bibitem{R10}
D. Arndt and B. C. Tiburzu, Phys.Rev {\bf D69} (2004) 014501

\bibitem{leinweber}
 D.B. Leinweber, T. Draper, R.M. Woloshyn, Phys.Rev. {\bf D48} (1993) 2230
 
\bibitem{R11}
E. Jenkins, X.-d. Ji and A.V. Manohar, Phys. Rev. Lett. {\bf 89} (2002) 242001

\bibitem{R12}
B. L. Ioffe and A. V. Smilga, Nucl. Phys. {\bf B232} (1984) 109

\bibitem{R13}
T. M. Aliev and M. Savci, Phys. Rev. {\bf D60} (1999) 114031

\bibitem{fxlee15} F. X. Lee, Phys.Rev. {\bf C57} (1998) 322

\bibitem{fxlee2} F. X. Lee and X.-Y. Lui Phys. Rev. {\bf D66} (2002) 014014

\bibitem{wyler} 
A.~Khodjamirian and D.~Wyler,
preprint hep-ph/0111249 (2001).

\bibitem{R14}
H.J. Lipkin, Phys. Rev. {\bf D7} (1973) 846

\bibitem{R15}
H. F. Jones and M.D. Scadron, Ann. Phys. {\bf 81} (1973) 1

\bibitem{R16}
R. C. E. Devenish, T.S. Eisenschitz and J. G. K{\"o}rner, Phys. Rev. {\bf D14}
(1976) 3063

\bibitem{zamir1} A.Ozpineci, S.B. Yakovlev, V.S. Zamiralov, preprint, hep-ph/0311271 (2003)

\bibitem{zamir2} A.Ozpineci, S.B. Yakovlev, V.S. Zamiralov, Mod.Phys.Lett. {\bf A20} (2005) 243 

\bibitem{ball:wave} P. Ball, V. M. Braun, N. Kivel, Nucl. Phys. {\bf B649} (2003) 263

\bibitem{hwang} W.-Y. P Hwang and K.-C. Yang, Phys. Rev. {bf D49} (1994) 460

\bibitem{fxlee1} F. X. Lee, Phys.Rev. {\bf D57} (1998) 1801

\bibitem{pdg} S. Eidelman {\it et al.}, Phys. Lett. {\bf B592} (2004) 1 

\bibitem{belyaev1} V. M. Belyaev and Ya. I. Kogan, Yad. Fiz. {\bf 40} (1984) 1035

\bibitem{belyaev2} V. M. Belyaev and B. L. Ioffe, JETP {\bf 56} (1982) 493

\bibitem{buchmann} A. J. Buchmann, E. Hernandez, and A. Faessler, Phys. Rev. {\bf C55} (1997) 448

\bibitem{selex} V.V. Molchanov {\em et al.} [SELEX Coll.] Phys. Let. {\bf B590} (2004) 161
\end{thebibliography}
\end{document}